%% file: acl_latex.tex
\pdfoutput=1

\documentclass[11pt]{article}

\usepackage[preprint]{acl}

\usepackage{times}
\usepackage{latexsym}
\usepackage{multirow}
\usepackage{booktabs}
\usepackage{tablefootnote}
\usepackage{arydshln}
\usepackage[normalem]{ulem}
\usepackage{graphicx}
\usepackage{listings}
\usepackage{colortbl}
\usepackage[T1]{fontenc}

\usepackage[utf8]{inputenc}
\usepackage{xspace}

\usepackage{microtype}

\usepackage{inconsolata}
\usepackage{algorithm}
\usepackage{algorithmic}
\usepackage{graphicx}
\usepackage{amsmath}
\usepackage{amssymb}
\usepackage{pgfplots}
\pgfplotsset{compat=1.18}
\usepackage{mdframed}
\usepackage{longtable}
\newcommand{\method}{\textsc{PaT}\xspace}
\title{\method: Parameter-Free Audio-Text Aligner to Boost Zero-Shot Audio Classification}

\author{
    Ashish Seth$^{1}$,
    Ramaneswaran Selvakumar$^{1}$,
    Sonal Kumar$^{1}$,
    Sreyan Ghosh$^{1}$,\\
    \bf Dinesh Manocha$^{1}$ \\
    $^{1}$University of Maryland, College Park\\
    \texttt{\{aseth125, ramans, sonalkum, sreyang, dmanocha\}@umd.edu}
}

\begin{document}
\maketitle
\begin{abstract}
Audio-Language Models (ALMs) have demonstrated remarkable performance in zero-shot audio classification. In this paper, we introduce \method (\textbf{P}arameter free \textbf{A}udio-\textbf{T}ext aligner), a simple and \textit{training-free} method aimed at boosting zero-shot audio classification performance of CLAP-like ALMs. To achieve this, we propose to improve the cross-modal interaction between audio and language modalities by enhancing the representations for both modalities using mutual feedback. Precisely, to enhance textual representations, we propose a prompt ensemble algorithm that automatically selects and combines the most relevant prompts from a datastore with a large pool of handcrafted prompts and weighs them according to their relevance to the audio. On the other hand, to enhance audio representations, we reweigh the frame-level audio features based on the enhanced textual information. Our proposed method does not require any additional modules or parameters and can be used with any existing CLAP-like ALM to improve zero-shot audio classification performance. We experiment across 18 diverse benchmark datasets and 6 ALMs and show that the \method outperforms vanilla zero-shot evaluation with significant margins of 0.42\%-27.0\%. Additionally, we demonstrate that \method maintains robust performance even when input audio is degraded by varying levels of noise. Our code will be open-sourced upon acceptance~\footnote{Code: \url{https://github.com/cs20s030/PAT.git}}.
\end{abstract}




\section{Introduction}

Advancements in multimodal language models (MLMs) have significantly improved performance across various modalities and applications~\cite{turian2022hear,yang2021superb}. Audio-Language Models~\citep{ghosh2024gama}, specifically Audio-Language Encoders (ALEs) like CLAP~\citep{guzhov2022audioclip,laionclap2023}, are a distinct type of MLMs that learn a shared representation space between the audio and language modalities. Trained on large-scale audio-caption pairs, these models acquire diverse audio concepts during pre-training, enabling them to generalize to new, unseen audio categories. This capability allows ALEs to excel in zero-shot audio classification, accurately categorizing any set of classes described with natural language during inference. Such flexibility is essential for ALEs to effectively adapt to dynamic environments with diverse and unknown sounds.


\begin{figure}[t]
    \centering
    \includegraphics[width=\columnwidth]{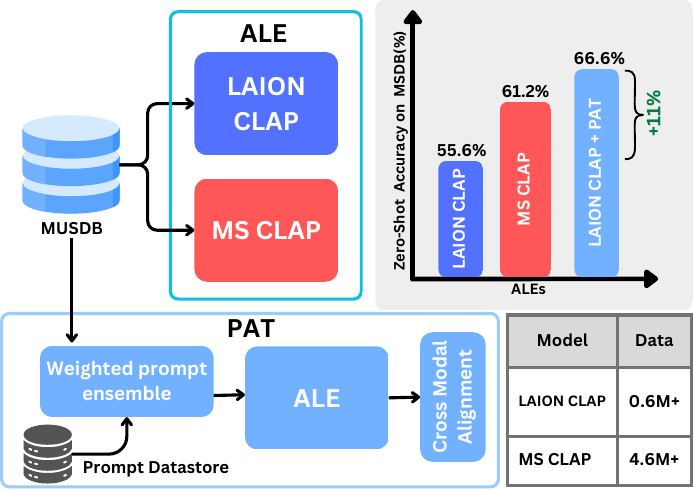}
    \caption{\small Comparison of zero-shot audio classification performance of ALEs (LAION CLAP~\citep{laionclap2023} and MS CLAP~\citep{elizalde2023clap}) with and without \method on MUSDB~\citep{Bertin-Mahieux2011} (music genre classification). Our proposed training-free method significantly enhances zero-shot performance, even in low-resource domains where the ALEs have limited training data.}
    \label{fig:pad_hero}
    \vspace{-1.5em}
\end{figure}


Significant progress has been made to enhance the zero-shot performance of ALEs across a wide range of audio classification tasks. For instance, Contrastive Language Audio Pretraining (CLAP)~\citep{laionclap2023} was one of the first ALE models to achieve notable zero-shot improvements by utilizing descriptive audio captions, as opposed to the class labels used in earlier models like Wav2CLIP\citep{wu2022wav2clip} and AudioCLIP~\citep{guzhov2022audioclip}. Additionally, ALEs such as LAION-CLAP~\citep{laionclap2023} and MS-CLAP~\citep{elizalde2023clap} employ vast collections of audio-text pairs, while models like CompA-CLAP~\citep{ghosh2023compa} leverage complex learning objectives during pretraining to show zero-shot performance gains across various audio classification tasks. While these models show zero-shot improvement, it comes at the additional cost of pre-training them with either more refined learning objectives or by improving both the quality and quantity of audio-text pairs. To overcome these challenges, researchers have attempted to improve ALEs' audio classification capabilities through parameter-efficient transfer learning. Inspired by prompt learning~\citep{brown2020language} and adapter fine-tuning approaches~\citep{houlsby2019parameter}, Treff Adapter~\citep{liang2023adapting} and Audio Prompt Learner~\citep{liang2023adapting} incorporate learnable prompts or lightweight adapters to adapt both textual and audio features. Although these methods result in performance improvements, they introduce learnable parameters and require additional training phases with few-shot labelled data. On the other hand, training-free methods can utilize the zero-shot capabilities of ALEs, making them more scalable and efficient for real-world applications which can contain diverse and unknown sounds. To the best of our knowledge, there exists no \textit{training-free} method to improve the zero-shot classification capabilities of such models.


\noindent{\textbf{Our Contribution.}} To this end, we propose \textbf{\method}: \textbf{P}arameter-free \textbf{A}udio \textbf{T}ext Aligner, a simple, parameter-free, and training-free approach for boosting zero-shot audio classification performance in ALEs. We first identify two major issues with current ALEs: 1) Current ALEs rely on simplistic prompts like ``The sound of <label>'' during zero-shot evaluation, leading to suboptimal performance. 
2) During zero-shot transfer, when we align audio and text representations through cosine-similarity, audio representations are typically average pooled, which results in information loss and reduced discriminative power~\citep{ruderman2018pooling,springenberg2014striving}. Inspired by our findings, we propose two novel components: \emph{1) Weighted Prompt Ensemble:} To enhance zero-shot performance for ALEs, we enrich the textual embeddings by creating a task-agnostic prompt datastore comprising 400 unique prompts. These prompts are specifically designed to reduce the distribution shift between the zero-shot setting and the training data, where audio captions typically contain more than a single-word description, unlike previous approaches. Each prompt is generated from handcrafted templates, capable of accommodating unique sound labels while ensuring semantic coherence. Moreover, we show that naively using all the prompts for various downstream applications may not consistently yield the desired performance improvements. To address this, we introduce \emph{weighted prompt ensemble}, a method which computes a score for each prompt using the ALEs' response as feedback without the need for additional labeled data or training. These scores are then used as weights to perform a weighted ensemble of the text embedding associated with a prompt and a class label. \emph{2) Cross Modal Aligner:} We introduce a \emph{cross-modal aligner} that enhances audio representations with text-guided information. Specifically, we first compute an attention map using parameter-free attention mechanisms between the frame-level audio representations and the enriched textual representations associated with each label. Next, we use this attention map to perform weighted pooling on the frame-level audio representations. Finally, we compute the cosine similarity between the enriched audio and text representations to perform zero-shot classification. To summarize, our main contributions are:
\begin{itemize}
    \setlength\itemsep{0em} 
    \item We propose \method, a novel approach to improve zero-shot audio classification performance in a \textit{training-free} fashion. \method introduces a cross-modal interaction approach aimed at improving audio-text alignment by enhancing both audio and textual representations in a zero-shot setting.
    \item We evaluate \method across multiple ALEs on 18 audio classification datasets and show that \method achieves 0.42\%-27.0\% improvement over our baselines.
    \item We further investigate \method's robustness to noisy audio to show that \method consistently outperforms our baselines under varied noise augmentation settings.
\end{itemize}

\begin{figure*}[t]
    \centering
    \includegraphics[trim=0 0.35cm 0 0,width=\textwidth]{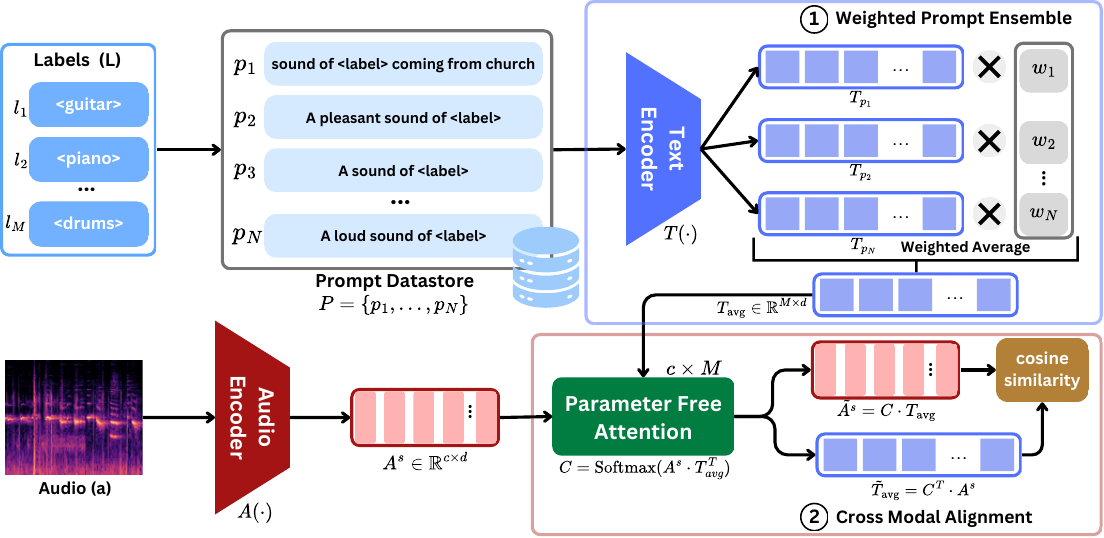}
    \caption{\small Illustration of \method. \method improves the zero-shot capabilities of ALEs by enriching audio-text representations in a parameter and training-free fashion. \method consists of two major components: \textcircled{\raisebox{-0.9pt}{1}} \emph{Weighted prompt ensemble} that first utilizes an in-house generic prompt datastore to transform class labels into diverse textual descriptions, which are then encoded by a text encoder. Further, each prompt is assigned a unique score based on the level of uncertainty it introduces during zero-shot prediction (less uncertainty results in a higher score). A weighted average is then performed to generate task-specific, semantically rich textual representations. \textcircled{\raisebox{-0.9pt}{2}} Next, the enriched textual representations are used to guide the enhancement of audio representations using a novel zero-shot \emph{cross model alignment}. Precisely, frame-level audio representations are paired with enhanced textual representations to compute a parameter-free attention map, which is used in performing audio and text-guided transformations. Finally, the transformed frame-level audio representations are pooled, and the audio-text-guided information is added to the original logit space, boosting the ALE's zero-shot prediction capabilities.}
    \label{fig:main_diag}
\end{figure*}

\section{Related Work}
\noindent{\textbf{Audio Language Encoders (ALEs).}} Previous explorations on developing multimodal encoders to learn shared representations across different modalities have shown significant promise. For example, building on contrastive pre-training techniques from vision-language models like CLIP~\citep{radford2021learning}, audio-language encoders (ALEs) have made advances in audio-language understanding, achieving state-of-the-art zero-shot performance across various audio classification tasks. Early efforts, such as Wav2Clip~\citep{wu2022wav2clip} and AudioCLIP~\citep{guzhov2022audioclip}, focused primarily on aligning audio representations with labels. More recent approaches, like CLAP~\citep{laionclap2023,elizalde2023clap} have shifted towards explicitly mapping audio representations to textual descriptions, leading to significant improvements in zero-shot performance across diverse audio classification tasks. However, most prior ALEs have relied on compute-intensive strategies, such as refining existing multimodal alignment objectives during pre-training~\citep{ghosh2023compa,kim2024enclap,ghosh2024reclap} or increasing the number of parameters and training data~\citep{elizalde2023clap}. In contrast, training-free approaches to improve zero-shot performance for ALEs remain largely underexplored.

\noindent{\textbf{Downstream Adaptation of ALEs for Audio Classification.}} Recent methods have employed various parameter-efficient transfer learning approaches, primarily inspired by prompt learning~\citep{brown2020language} and adapter fine-tuning~\citep{houlsby2019parameter}. For instance, Treff Adapters~\citep{liang2023adapting} introduce an additional learnable cross-attention module to enhance cross-modal interaction in few-shot settings. Other approaches, such as Audio Prompt Learners~\cite{10446472} and PALM~\cite{hanif2024palm}, use soft prompting to append learnable prompt tokens to textual representations. Although these methods show improvements in audio classification tasks, they fail to leverage the core advantage of ALEs, which is zero-shot transfer.
\section{Preliminaries}
\subsection{Zero-shot transfer in ALEs}
Let $D$ be a downstream audio classification dataset with $M$ class labels, $L \leftarrow \{l_1, \dots, l_M\}$. First, we insert these labels into a fixed prompt template $p$, such as ``The sound of <label>,'' where the ``<label>'' token is replaced with each unique class label. This creates $M$ textual inputs: $p(L) \leftarrow {p(l_1), \dots, p(l_M)}$. Next, we use text encoders $T(\cdot)$ to transform these textual inputs into latent representations: $T_p \leftarrow {T_p(l_1), \dots, T_p(l_M)}$. Similarly, for each audio input $a_i$, we use audio encoders $A(\cdot)$ to obtain frame-level audio representations $A^s_i \in \mathcal{R}^{c \times d}$, which we then pool to produce compact representations $A_i \in \mathcal{R}^{1 \times d}$. To ensure consistent dimensionality between audio and text representations, we pass both through fully connected layers of the same size. Finally, we compare the features using cosine similarity to generate the classification logits: $\mathcal{L}_{\text{pred}}(a_i) \leftarrow A_i \cdot T_{p}^T$
\subsection{Prompt Ensemble}
Let $P\leftarrow \{p_1,...,p_N\}$ be a collection of unique prompts. We first encode each prompt $p_i$ with class labels $l_j$ using a text encoder and then average the textual representations of similar labels: $T_{\text{avg}} \leftarrow \{\sum_{i=i}^N T_{p_i}(l_1)\},...,\sum_{i=i}^N T_{p_i}(l_m)\}$. We recognize that one key challenge in applying prompt ensembles to Audio-Language Encoders (ALEs) is the lack of a collection of diverse prompts specifically tailored for zero-shot audio classification.

\begin{figure}[t]
    \centering
    \begin{mdframed}[linewidth=1pt, linecolor=black, leftmargin=1pt, rightmargin=1pt, innerleftmargin=10pt, innerrightmargin=10pt, innertopmargin=4pt, innerbottommargin=2pt, backgroundcolor=gray!20, roundcorner=5pt]
        \textit{\textbf{(1) Prompt Template:}} A \textit{<attribute>} sound of <label>

        \noindent\textit{\textbf{Attribute:}} ``loud'', ``pleasant'', etc.

        \noindent\textit{\textbf{Example:}} ``A loud sound of a car'', ``a pleasant sound of a guitar.''

        \noindent\textit{\textbf{Paraphrased:}} ``A car produces a loud sound.'', ``A guitar emits a pleasant sound.''

        \vspace{0.2cm}

        \textit{\textbf{(2) Prompt Template:}} A sound of <label> coming from <source>

        \noindent\textit{\textbf{source:}} ``church'', ``garden'', etc.

        \noindent\textit{\textbf{Example:}} ``A sound of bells coming from a church'', ``a sound of a cat coming from a garden''.

        \noindent\textit{\textbf{Paraphrased:}} ``Bells ringing can be heard from the church.'', ``The sound of a cat is coming from the garden.''
    \end{mdframed}
    \caption{Examples of prompt templates and their paraphrased versions used in the prompt datastore.}
    \label{fig:prompt_templates}
\end{figure}

\section{Methodology}
Fig.\ref{fig:main_diag} demonstrates our proposed method \method. We propose two simple training-free extensions for existing zero-shot transfer learning in ALEs. To enhance textual representation, we introduce a weighted prompt ensemble that selectively identifies and scores prompts which are more relevant to an unseen downstream task. To enhance audio representation, we introduce a cross-modal alignment that utilizes a parameter-free attention mechanism to align frame-level audio representations with textual descriptions in a zero-shot setting. In upcoming subsections, we explain \method in detail.









\subsection{Prompt Datastore}
We adopt prompt ensemble for zero-shot audio classification tasks by developing a robust prompt datastore for ALEs. Specifically, our prompt datastore consists of 400 semantically and linguistically diverse prompts. These prompts are designed to minimize the distribution shift between the zero-shot setting and training data, where audio captions typically contain more than a single-word description. We first use GPT-4o~\citep{openai2024gpt4technicalreport} to generate 600 prompts by providing diverse templates and examples to guide generation. To ensure semantic diversity in our prompts, we design a variety of prompt-generation templates to avoid vague or overly open-ended prompts, such as "How does the sound <label> occur?". Furthermore, to ensure linguistic diversity, we ask GPT-4o to generate a paraphrased version of the initial set of generated prompts. Finally, we conduct a manual filtering process, selecting 400 prompts by discarding those that are inaccurate or repetitive. Fig.~\ref{fig:prompt_templates} shows a few examples of prompt templates (See Appendix~\ref{tab:prompt_score} for more examples of prompts generated).

\subsection{Weighted Prompt Ensemble}
\noindent{\textbf{Motivation.}} 
Through our experiments, we discover that despite the vanilla prompt ensemble method consistently outperforming single prompts like ``The sound of <label>'', not all prompts in a large prompt datastore $P$ are equally relevant for unseen downstream tasks. To address this in VLEs, prior work has utilized prompt engineering to create task-specific prompts by analyzing output labels for unseen tasks~\citep{zhou2022learning}. However, we argue that such an approach requires additional manual effort and is not scalable across diverse downstream tasks.

In response to this, we propose a Weighted Prompt Ensemble (WPE) algorithm in \method to adapt prompt ensembles across unseen downstream tasks by scoring individual prompts automatically without requiring additional parameters or training. Our approach is guided by a simple intuition: \emph{Given a set of audio data and class labels, prompts that introduce greater uncertainty in label predictions should receive lower scores, and vice versa}. In line with the prior works~\citep{zhang2023decoupling}, we quantify uncertainty in model predictions using \emph{max prediction logits}. As shown in Fig.~\ref{fig:main_diag}, for a given prompt $p$ in prompt datastore $P$, audio representation $A$, and textual representation $T_p \in \mathbb{R}^{M\times d}$, where $M$, $d$ are the number of class labels and embedding dimensions respectively, we first compute the prediction logits $\mathcal{L}_{p} \leftarrow A \cdot T_{p}^T$. The prompt score $w_p$ is then determined by calculating a cumulative sum of maximum prediction logits. We formally define this as:
\begin{equation}
    w_p = \sum_{i=1}^{|D|} \text{max}_j \mathcal{L}_\text{pred}[i,j], \forall j \in \{1, \dots, M\}
\end{equation}
where $|D|$ is the number of samples in an unseen downstream task $D$. Finally, we apply softmax normalization to the prompt scores and perform a weighted averaging of the prompt-specific textual representations to obtain enriched textual representation $T_{\text{avg}}$, as shown below:
\begin{equation}
    T_{\text{avg}} = \sum_{p=1}^{N} w_p * T_p
\end{equation}
where $N$ is the number of unique prompts. We summarize the complete process of our weighted prompt ensemble in Algorithm~\ref{alg:wpe}

\input{algorithm}

\subsection{Cross-Modal Alignment}
\noindent{\textbf{Motivation.}} We observe that ALEs encode audio and text modalities independently, preventing them from utilizing cross-modal interactions before making final predictions. Additionally, audio representations are typically average-pooled, leading to a loss of fine-grained information in the audio.

Inspired by our findings, we propose parameter-free cross-modal alignment in \method to improve the audio-text alignment in ALEs in a zero-shot fashion. As shown in Fig.~\ref{fig:main_diag}, we utilize the intermediate frame-level audio representations $A^s \in \mathbb{R}^{c\times d}$ and enriched textual representations $T_{avg} \in \mathbb{R}^{M\times d}$ to compute parameter free attention weights $c_{i,j} \in C$ that denotes the correlation between individual audio frames and class labels. We formulate this below:
\begin{equation}
    C = A^s \cdot T^T_{avg} \in \mathbb{R}^{c\times M} 
\end{equation}
Next, we utilize our parameter-free attention weights $C$ to enhance cross-modal alignment between the audio and textual representations. To achieve this, we project the original audio representation $A^s$ and the average textual representation $T_{\text{avg}}$ using $C$ as follows:

\begin{equation}
    \Tilde{A^s} = \text{Softmax}(C) \cdot T_{\text{avg}} \in \mathbb{R}^{c \times d}
\end{equation}
\begin{equation}
    \Tilde{T}_{\text{avg}} = \text{Softmax}(C^T) \cdot A^s \in \mathbb{R}^{M \times d}
\end{equation}
where $\Tilde{A^s}$ represents the audio-guided representations that amplify the information of labels strongly associated with the audio, and $\Tilde{T}_{\text{avg}}$ represents the text-guided representations that enhance the information of the audio frames strongly correlated with specific labels. We then pool the transformed frame-level audio representation $\Tilde{A^s}$ as follows: $\Tilde{A} \leftarrow \text{AvgPool}(\Tilde{A^s})$. The transformed representations $\Tilde{A}$ and $\Tilde{T}_{\text{avg}}$ are then used to create audio-guided and text guide logits, $\mathcal{L}_{\text{audio}}\leftarrow \Tilde{A}\cdot T_{\text{avg}}^T$ and $\mathcal{L}_{\text{text}}\leftarrow A \cdot \Tilde{T}_{\text{avg}}^T$ respectively. Finally, we combine the audio and text-guided logits with the original logits $\mathcal{L}_{\text{pred}} \leftarrow A \cdot T^T_{\text{avg}}$. We formulate this below:
\begin{equation}
    \mathcal{L}_{\text{combine}} = \mathcal{L}_{\text{pred}} + \beta_1 \mathcal{L}_{\text{audio}} + \beta_2 \mathcal{L}_{\text{text}}  
\end{equation}
where $\beta_1$ and $\beta_2$ are the hyper-parameter weights for audio and text-guided logits, respectively. Additionally, we provide details on hyper-parameter tuning in Appendix~\ref{sec:hyperparameter}

\input{main_results_table}

\section{Experimental Setup}
\subsection{Evaluation Datasets and Metric}
For zero-shot evaluation, we utilized 18 open-source audio classification datasets covering a broad range of musical and non-verbal audio types. Specifically, for music-related downstream tasks such as instrument and genre classification, we present results on eight widely-used musical datasets, including NSynth (NSynth Inst/Src)~\citep{engel2017neural}, Beijing Opera (Beijing Op)~\citep{bejingopera}, MedleyDB~\citep{bittner2014medleydb}, MUSDB~\citep{musdb18}, GTZAN~\citep{tzanetakis_essl_cook_2001}, and Mridangam Stroke/Tonic (Mri. St/ Mri. To)~\citep{turian2022hear}. Additionally, we report results on non-verbal sound classification tasks using datasets like ESC-50~\citep{piczak2015dataset}, UrbanSound8K (USD8K)\citep{Salamon:UrbanSound:ACMMM:14}, TUT-Urban (TUT)\citep{Mesaros2018_DCASE}, VocalSound (VS)\citep{gong_vocalsound}, SESA\citep{spadini2019sound}, CochlScene~\citep{jeong2022cochlscene}, DCASE Task 4~\citep{mesaros2017dcase}, and Gunshot Triangulation (GT)~\citep{turian2022hear}. We also conduct an evaluation on multi-label datasets like FSD50K~\citep{fonseca2021fsd50k} and AudioSet~\citep{gemmeke2017audio}. For zero-shot evaluation, we report the mean average precision (mAP) for AudioSet and FSD50K and accuracy for the other datasets, which averaged over 5 runs. Appendix~\ref{sec:dataset_details} provides additional dataset details.
\subsection{Baselines}
We demonstrate the scalability and robustness of the \method across five open-source audio-language encoders (ALEs). Specifically, we integrate \method with Wav2CLIP~\citep{wu2022wav2clip}, trained with 200k audio-label pairs and various CLAP models, including LAION CLAP (L-CLAP)~\citep{laionclap2023}, trained on over 633K audio-text pairs; CompA CLAP, which is further trained on 60K compositionally aware audio-text pairs; LAION CLAP MUSIC (LM-CLAP), which is further trained on music datasets; MS CLAP-22~\citep{CLAP2022}, trained on 128K+ audio-text pairs; and MS CLAP-23~\citep{elizalde2023clap}, which is trained on 4.6M pairs covering diverse audio types such as speech, music, and non-verbal sounds. We re-evaluate the zero-shot accuracy of publicly available ALEs under similar compute settings using A6000 GPUs.

\subsection{Audio Augmentations}
We further evaluate \method's robustness in noisy settings by augmenting audio with various kinds of audio augmentations. Specifically, we apply \textit{gaussian noise} to simulate real-world background noise, \textit{pitch shift} to test frequency variation, and \textit{polarity inversion} to check the model's sensitivity to phase changes. Additionally, we use \textit{delay} to introduce echo effects, \textit{gain} to assess performance under varying volume levels, and both \textit{low pass} and \textit{high pass filters} to evaluate the \method ability to handle reduced frequency ranges. Lastly, we use \textit{reverb} to simulate different reverberant acoustic environments. Appendix~\ref{sec:noise_augs} provides additional details on various audio augmentations.
\input{ablation_method_table}
\input{cma_analysis_table}

\input{prompt_number}
\input{prompt_score}
\section{Results and Result Analysis}
\label{sec:results_analysis}
\subsection{Main Results}
We summarize the results of \method applied to 6 different ALEs across 16 audio classification datasets, including 8 non-verbal speech and non-speech sound datasets and 8 musical datasets, in Table~\ref{tab:main_result}. Our key finding can be summarized as follows: 1) \method consistently outperforms the vanilla zero-shot (ZS) approach across all baselines and datasets, achieving an absolute improvement of 0.42\%-27\%. This underlines \method strength to generalize across diverse audio classification tasks and scale with different ALEs in a training-free fashion. 2) \method shows a remarkable performance boost even for less-pretrained ALEs. For instance, LAION-CLAP~\citep{laionclap2023} pre-trained on 0.6M audio-text pairs, gains an 11\% boost on MUSDB~\citep{rafii2017musdb18} when combined with \method, even surpassing MS CLAP~\citep{elizalde2023clap} by 5\%, which has been pre-trained on 4.6M pairs. We attribute such gains to our cross-modal aligner that utilizes enriched textual representations to reweigh audio representations. 3) \method shows a slight performance degradation on ALEs like Wav2CLIP~\citep{wu2022wav2clip}. As mentioned by~\citep{elizalde2023clap}, these models are pre-trained solely on audio labels rather than textual descriptions, thus limiting them to interpret prompts.
\input{noise_table}

\subsection{Ablation on various \method Components}
We conduct extensive ablations on individual \method components, as shown in Table~\ref{tab:ms_clap_results} and Table~\ref{tab:cma}. For all ablation studies, we use the best-performing zero-shot model, MS-CLAP, unless stated otherwise. In Table~\ref{tab:ms_clap_results}, we compare the performance gains from using Prompt Ensemble (PE) versus Weighted Prompt Ensemble (WPE), which is employed in \method, averaging the zero-shot accuracy across all datasets. Overall, both PE and WPE outperform the baseline, with WPE consistently outperforming PE even with or without Cross-Modal Alignment (CMA). With this, we show that selectively scoring prompts that reduce uncertainty in model predictions, as opposed to using uniform scoring, leads to improved zero-shot performance. Table~\ref{tab:cma} further explores CMA by ablating different combinations of original, audio-, and text-guided logits. We show that combining all the logits gives the biggest improvement in zero-shot performance.   
\subsection{Ablation on Number of Prompts}
Fig.~\ref{fig:wer_comparison} shows the effect of increasing the number of prompts when applying the weighted prompt ensemble in \method. The key findings are: 1) While adding more prompts to the prompt datastore improves zero-shot performance in \method, the performance of MSCLAP-23 plateaus after reaching 400 prompts. This suggests that naively increasing the number of prompts may not lead to further gains in zero-shot performance for audio classification tasks. 2) The Weighted Prompt Ensemble (WPE) used in \method consistently surpasses the standard Prompt Ensemble (PE), even with fewer prompts. Table~\ref{tab:prompt_score} lists the top 2 highest-scoring prompts generated by MSCLAP-23 using \method on both sound datasets (e.g., ESC-50) and music datasets (e.g., NSynth Inst). We observe that, for sound datasets, prompts associating sounds to random sources like ``parking lot'' or ``cliff edge'' receive higher scores. In contrast, for music datasets, prompts describing random audio attributes like ``minimal'' or ``loud'' tend to score higher. We extend this analysis in Appendix~\ref{sec:prompt_scores}, where we show the top 10 best-scoring prompts for each downstream task.
\subsection{Zero-Shot Evaluation under Noisy Setting}
Table~\ref{tab:noise table} shows the performance improvement of MSCLAP-23 using \method across 16 audio classification tasks under noisy conditions, compared to the baseline zero-shot scores. In particular, individual audio samples in each task are subjected to five different types of augmentation, which modify key audio features such as pitch, frequency, amplitude, and phase. Under noisy settings, \method is able to show promising results by achieving an absolute improvement of 0.10-11.15\% when compared with simple zero-shot evaluation. This highlights the robustness of \method to classify out-of-distribution audio samples without the need for additional few-shot data or training. Additionally, we provide results on 3 other audio augmentations: delay, low pass and reverb in Appendix~\ref{sec:noise_augs}
\section{Conclusion}
We introduce \method, a parameter- and training-free approach to improve zero-shot audio classification for audio-language encoders (ALEs). Our studies indicate that current methods that guarantee audio classification improvements are not capable of leveraging the core strength of ALEs, which is zero-shot transfer and often require extensive training or additional parameters to adapt audio and textual representations for an unseen task. To overcome this limitation, \method offers two key contributions: 1) a novel prompt scoring method, called weight prompt ensemble, which adapts prompt ensembling for unseen audio classification tasks in a training-free fashion, and 2) a cross-modal alignment framework, which uses parameter-free attention to better align audio and textual representation. \method significantly boosts zero-shot performance in ALEs across various audio classification datasets.

\section{Limitation and Future Work}
In this section, we highlight a few limitations and potential future direction of our proposed method, \method 

\begin{itemize}
    \item Due to compute constraints, we did not evaluate \method across speech classification tasks. In future, we plan to release \method evaluation scores against various speech-related downstream tasks such as Keyword spotting, Emotion Recognition, etc.
    \item We acknowledge that \method increases the space complexity of existing zero-shot evaluation algorithms by introducing a prompt datastore.
    \item While \method shows significant improvements, it is limited by ALE's existing knowledge space to classify unknown sounds. In future, we plan to integrate parameter-efficient methods such as soft prompting to make \method robust towards the new evolving sounds.
\end{itemize}

\bibliography{custom}

\appendix
\section{Appendix}
\label{sec:appendix}
In the appendix, we provide:
\begin{enumerate}
    \item Section~\ref{sec:baselines_details}: Baseline Details
    \item Section~\ref{sec:dataset_details}: Dataset Details
    \item Section~\ref{sec:noise_augs}: Audio Augmentations
    \item Section~\ref{sec:hyperparameter}: Hyper-Parameter tuning
    \item Section~\ref{sec:other_details}: Additional Details
    \item Section~\ref{sec:prompt_scores}: Prompt Scores
\end{enumerate}

\section{Baseline Details}
\label{sec:baselines_details}
\noindent\textbf{LAION-CLAP}~\citep{laionclap2023}. This is a contrastive language-audio pretraining (CLAP) model from LAION-AI trained on LAION-Audio-630K~\cite{laionclap2023}, a large collection of 633,526 audio-text pairs from different data sources. To improve the model's ability to handle audio inputs of variable lengths and boost overall performance, it integrates a feature fusion mechanism and keyword-to-caption augmentation. This enables the model to effectively align and process both audio and text data for enhanced learning.

\noindent\textbf{LAION-CLAP Music}~\citep{laionclap2023}. This is a music-specific version of the LAION-CLAP model. This version is trained both on audio and music, with the LAION-Audio-630K dataset contributing a major portion of its training data. The details of the music-text data being used for training are not specified.

\noindent\textbf{MS-CLAP 22}~\citep{CLAP2022}. This is a contrastive language-audio pretraining (CLAP) model from Microsoft. This version is trained on 128k audio and text pairs. 

\noindent\textbf{MS-CLAP 23}~\citep{elizalde2023clap}. This is a follow-up to the MS-CLAP 22, from Microsoft. This version of CLAP uses two innovative encoders and is trained on massive 4.6M audio-text pairs.  To learn audio
representations, the authors trained an audio encoder on 22 audio tasks instead of the standard training of sound event classification. To learn language representations, they trained an autoregressive decoder-only model instead of the standard encoder-only models.

\noindent\textbf{Wav2CLIP}~\citep{wu2022wav2clip}. This is a robust audio representation learning method by distilling from Contrastive Language-Image Pre-training (CLIP)~\citep{radford2021learning}. Wav2CLIP is a model that maps audio into a shared embedding space alongside images and text, enabling multimodal tasks like zero-shot classification and cross-modal retrieval. It achieves competitive performance on downstream tasks with only about 10\% of the data needed by fully supervised models. Additionally, Wav2CLIP is more efficient in pre-training, as it focuses solely on the audio modality and does not require joint training of visual and auditory models, unlike some competing methods.

\noindent\textbf{CompA}~\citep{ghosh2023compa}. This is a CLAP model that is trained specifically to enhance its compositional reasoning abilities. The authors introduce improvements to contrastive training by incorporating composition-aware hard negatives, allowing for more precise and focused training. Additionally, they propose a modular contrastive loss designed to help the model learn fine-grained compositional understanding.

\section{Dataset Details}
\label{sec:dataset_details}
\noindent\textbf{ESC-50:}\footnote{\url{https://github.com/karolpiczak/ESC-50}}~\citep{piczak2015dataset} The ESC-50 dataset is a labelled collection of 2,000 environmental audio recordings, every 5 seconds in length. The dataset is designed for sound classification tasks and contains recordings organized into 50 semantically distinct classes, with 40 examples per class. These classes are further grouped into 5 major categories, which include Animals, Natural soundscapes \& water sounds, Humans, non-speech sounds, Interior/domestic sounds, and Exterior/urban noises.

\noindent\textbf{USD-8K:}\footnote{\url{https://urbansounddataset.weebly.com/urbansound8k.html}}~\citep{Salamon:UrbanSound:ACMMM:14} The UrbanSound8K dataset is an audio collection that contains 8,732 labelled sound excerpts, each with a duration of up to 4 seconds. The dataset is designed to represent various urban sound environments, with recordings organized into 10 distinct classes: air\_conditioner, car\_horn, children\_playing, dog\_bark, drilling, engine\_idling, gun\_shot, jackhammer, siren, and street\_music. These classes are derived from the urban sound taxonomy

\noindent\textbf{TUT Urban Acoustic Scenes (TUT):}\footnote{\url{https://zenodo.org/records/2589280}}~\citep{mesaros2017tut} The TUT Acoustic Scenes 2019 is a large-scale collection of environmental recordings from various urban sound environments. It consists of 10 classes, each representing a different acoustic scene, such as Airport, Metro station, Park, and Residential area. Each recording is 10 seconds long.

\noindent\textbf{Vocal Sound (VS):}\footnote{\url{https://github.com/YuanGongND/vocalsound}}~\citep{gong_vocalsound} The VocalSound dataset consists of 21,024 crowdsourced recordings representing 6 classes of vocal sounds: laughter, sighs, coughs, throat clearing, sneezes, and sniffs. These recordings were collected from 3,365 unique subjects, providing a diverse set of vocal sound samples. This dataset is primarily used for the study and classification of non-verbal vocal sounds.

\noindent\textbf{DCASE Task 4:}\footnote{\url{https://dcase.community/challenge2017/}}~\citep{mesaros2017dcase} The DCASE Task 4 dataset is a sound event detection dataset with heterogeneous data and missing labels. It comprises two primary datasets: DESED and MAESTRO. DESED consists of 10-second-long audio clips that are either recorded in domestic environments or synthesized to simulate such environments. These recordings contain annotated sound events from 10 different classes. MAESTRO provides audio recordings with multiple temporally strong annotated events featuring soft labels across 17 classes. However, for the purposes of DCASE Task 4, only 11 classes from MAESTRO are considered.

\noindent\textbf{Gunshot Triangulation:}\footnote{\url{https://hearbenchmark.com/hear-tasks.html}}~\citep{turian2022hear} The Gunshot Triangulation dataset is designed for a novel multiclass classification task involving gunshot recordings. The dataset consists of 88 audio clips representing 22 gunshots from 7 different firearms, recorded in an open field using iPod Touch devices (Cooper and Shaw, 2020). The objective of the task is to classify the audio based on the specific iPod Touch that was recorded, effectively identifying the location of the microphone during each gunshot event.

\noindent\textbf{SESA:}\footnote{\url{https://zenodo.org/records/3519845}}~\citep{spadini2019sound} The Sound Events for Surveillance Applications (SESA) dataset consists of audio recordings sourced from Freesound. The audio files have durations of up to 33 seconds and are categorized into 4 classes: Casual (not a threat), Gunshot, Explosion, and Siren. This dataset is intended for tasks related to sound event detection and classification in surveillance contexts.

\noindent\textbf{Cochlscene:}\footnote{\url{https://github.com/cochlearai/cochlscene}}~\citep{jeong2022cochlscene}The CochlScene dataset is a crowdsourced collection consisting of 76,000 audio samples from 831 participants, covering 13 different acoustic scenes. Each audio file is 10 seconds in length. This dataset is used for the classification of diverse acoustic environments, providing a wide range of real-world audio scenes.

\noindent\textbf{Beijing Opera:}\footnote{\url{https://compmusic.upf.edu/bo-perc-dataset}}~\citep{bejingopera} The Beijing Opera Percussion Instrument dataset is a collection of audio examples featuring individual strokes from the four main percussion instruments used in Beijing Opera. These instrument classes include Bangu, Naobo, Daluo, and Xiaoluo. The dataset is designed for the study and classification of traditional Chinese percussion instruments within the context of Beijing Opera.

\noindent\textbf{GTZAN:}\footnote{\url{https://paperswithcode.com/dataset/gtzan}}~\citep{tzanetakis_essl_cook_2001} The GTZAN Genre dataset is widely used for music genre classification tasks. It contains 1,000 audio tracks, each with a duration of 30 seconds, categorized into 10 distinct genres: Blues, Classical, Country, Disco, Hip-hop, Jazz, Metal, Pop, Reggae, and Rock. Each genre is represented by 100 tracks.

\noindent\textbf{MUSDB:}\footnote{\url{https://sigsep.github.io/datasets/musdb.html}}~\citep{musdb18} The MUSDB18 dataset consists of 150 full-length music tracks, totalling approximately 10 hours of audio, covering various genres. It is primarily used for music source separation tasks. The dataset provides four stem labels: drums, bass, vocals, and others (which include all other non-specific instruments and sounds). Each track is broken down into these stems, allowing for detailed analysis and separation of musical components.

\noindent\textbf{Medley:}\footnote{\url{https://medleydb.weebly.com/}}~\citep{bittner2014medleydb} MedleyDB, is a dataset of annotated, royalty-free multitrack recordings. It was curated primarily to support research on melody extraction. For each song melody, f\textsubscript{0} annotations are provided, as well as instrument activations for evaluating automatic instrument recognition. The original dataset consists of 122 multitrack songs, out of which 108 include melody annotations.

\noindent\textbf{Mridangam Stroke:}\footnote{\url{https://compmusic.upf.edu/mridangam-stroke-dataset}}~\citep{anantapadmanabhan2014mridangam} The Mridangam Stroke dataset is a collection of 7,162 audio examples featuring individual strokes of the Mridangam, a pitched percussion instrument used in Carnatic music, a sub-genre of Indian classical music. The dataset captures strokes in various tonics and includes 10 different stroke types played on Mridangams. 

\noindent\textbf{Mridangam Tonic:}~\citep{anantapadmanabhan2014mridangam} This dataset is a subset of the Mridangam Stroke dataset. It includes 6 different tonic values associated with the 10 different stroke types played on the Mridangam.

\noindent\textbf{NSynth Instrument:}\footnote{\url{https://magenta.tensorflow.org/datasets/nsynth}}~\citep{nsynth2017} NSynth Instrument is a part of the NSynth audio dataset, which consists of 305,979 musical notes, each characterized by unique combinations of pitch, timbre, and envelope. The NSynth Instrument dataset focuses on the task of identifying the high-level instrument family to which each note belongs. These instrument families include categories like keyboard, string, brass, reed, and mallet. 

\noindent\textbf{NSynth Source:}~\citep{nsynth2017} NSynth Source is a subset of the NSynth dataset, where the task is to identify the method of sound production for each instrument's note. There are three categories of sound production: acoustic, electronic, and synthetic. Acoustic and electronic labels correspond to instruments recorded from physical sources, while the synthetic label applies to notes generated using digital synthesis. 

\section{Audio Augmentations}
\label{sec:noise_augs}
We use 8 types of audio augmentations in our experiments. We employ torchaudio-augmentations'~\footnote{\url{https://github.com/Spijkervet/torchaudio-augmentations/}} implementation to augment the audios.

\noindent{\textbf{Gaussian Noise.}} Gaussian noise augmentation is a technique where random noise, following a Gaussian (normal) distribution, is added to audio signals. This simulates real-world background noise and helps improve the robustness of audio models by exposing them to a variety of noisy conditions during training. In our experiment, we set the Minimum Signal-to-Noise Ratio to 0.0001 and the Maximum Signal-to-Noise Ratio to 0.01.

\noindent{\textbf{Pitch Shift.}} Pitch shift augmentation is an augmentation technique which involves changing the pitch of an audio signal without affecting its tempo. In our experiment, we set the minimum pitch shift to -7.0 semitones (downward shift) and the maximum pitch shift to 7.0 semitones (upward shift).

\noindent{\textbf{Polarity Inversion.}} Polarity inversion is an audio augmentation technique where the polarity of the audio waveform is inverted by multiplying the signal by -1. This flips the waveform vertically, turning positive amplitude values into negative ones and vice versa. Although it doesn't affect the audible characteristics of the sound to human listeners, it can be helpful in testing how models perform phase changes in audio signals.

\noindent{\textbf{Gain.}} Gain augmentation involves adjusting the amplitude of an audio signal by applying a gain factor, effectively changing the volume of the audio. It helps in checking how the models perceive the same audio with different volume levels. For our experiment, we set the minimum gain to -20 dB and the maximum gain to -1 dB. 

\noindent{\textbf{High Pass Filter.}} High-pass filter augmentation involves applying a filter that allows frequencies above a certain cutoff frequency to pass through while attenuating frequencies below that cutoff. In our experiment, we set the minimum cutoff frequency to 200 Hz and the maximum cutoff frequency to 1200 Hz.

\noindent{\textbf{Low Pass Filter.}} Low-pass filter augmentation involves applying a filter that allows frequencies below a certain cutoff frequency to pass through while attenuating frequencies above that cutoff.  In our experiment, we set the minimum cutoff frequency to 2200 Hz and the maximum cutoff frequency to 4000 Hz.

\noindent{\textbf{Delay.}} Delay augmentation involves adding a delayed version of the audio signal back onto itself, creating an echo effect. In our experiment, we set the volume factor to 0.5, minimum delay to 200 ms, maximum delay to 500 ms and delay interval to 50 ms.

\noindent{\textbf{Reverb.}} Reverb augmentation involves adding reverberation effects to audio signals, simulating the natural reflections of sound in an acoustic environment like a room or hall. In our experiment, we set the minimum reverberance to 0, maximum reverberance to 100, damping factor to 75 and room size to 100.

\section{Hyper-parameter tuning}
\label{sec:hyperparameter}
Table~\ref{tab:hp} shows the hyper-parameter tuning for audio-guided and text-guided logit weights, $\beta_1$ and $\beta_2$ in \method. Table~\ref{tab:best_hp} shows the list of best-performing hyper-parameters for \method when applied to MSCLAP-23 across all the evaluation datasets

\begin{table}[h]
    \centering
    \resizebox{\columnwidth}{!}{
    \begin{tabular}{@{}c|ccccc@{}} 
    \hline \hline
$(\beta_1,\beta_2)$     & (0.01, 0.1) & (0.05, 0.5) & (0.1, 0.02) & (0.01, 0.01) & (0.5, 0.5) \\ \hline
Nsynth Inst. & 64.34 & 62.12  & 65.10  & \textbf{66.34}  & 61.23 \\ \hline
ESC-50 & \textbf{94.80} & 92.13  & 93.21  & 91.89  & 93.08 \\ \hline
    \end{tabular}}
    \caption{\small The hyper-parameter tuning for \method applied on MSCLAP-23.}
    \label{tab:hp}
\end{table}

\begin{table}[h]
    \centering
    \resizebox{1.0\columnwidth}{!}{
    \begin{tabular}{@{}c|cc@{}} 
    \hline \hline
Dataset & $\beta_1$ (Audio Guided)& $\beta_2$ (Text Guided)\\ \hline
ESC-50 & 0.01 & 0.1\\ \hline
USD8K & 0.02 & 1.2\\ \hline
TUT & 0.18 & 2.2\\ \hline
VS & 0.04 & 0.5\\ \hline
DCASE & 0.11 & 3.1\\ \hline
Gunshot Tri & 0.01 & 0.13\\ \hline
SESA & 0.01 & 0.11\\ \hline
AudioSet & 0.01 & 0.01\\ \hline
FSD50K & 0.02 & 0.01\\ \hline
Cochlscene & 0.3 & 2.5\\ \hline
Beijing Op. & 0.13 & 1.8\\ \hline
GTZAN & 0.4 & 3.2\\ \hline
MUSDB & 0.02 & 2.5\\ \hline
Medley & 0.04 & 1.3\\ \hline
Mri. St. & 0.23 & 1.4\\ \hline
Mri. To. & 0.1 & 0.1\\ \hline
NSynth Inst & 0.01 & 0.01\\ \hline
NSynth Src & 0.42 & 2.13\\ \hline
    \end{tabular}}
    \caption{\small Best hyper-parameter for \method when used with MSCLAP-23 across 18 datasets}
    \label{tab:best_hp}
\end{table}

\section{Additional Details}
\label{sec:other_details}

\noindent\textbf{Model Parameters:} Among the ALEs that we use, LAION-CLAP and LAION-CLAP Music have $\approx$ 158M parameters. MSCLAP-22 has $\approx$ 196M parameters and MSCLAP-23 has $\approx$ 159M parameters. Wav2CLIP has $\approx$ 140M parameters. CompA has $\approx$ 300M parameters.

\noindent\textbf{Compute Infrastructure:} All our experiments are conducted on one NVIDIA A6000 GPUs. No training is required, and depending on the downstream task, a single inference run on a benchmark requires anywhere between 1 and 5 minutes. 

\noindent\textbf{Implementation Software and Packages:} For our baselines, we use the original
GitHub repository provided by the authors: LAION-CLAP~\footnote{\url{https://github.com/LAION-AI/CLAP/tree/main}}, CompA-CLAP~\footnote{\url{https://github.com/Sreyan88/CompA}}, MS-CLAP~\footnote{\url{https://github.com/microsoft/CLAP/tree/main}}, Wav2CLIP~\footnote{\url{https://github.com/descriptinc/lyrebird-wav2clip}}. 

\noindent\textbf{Potential Risks:} To create the prompt datastore, we use GPT4o, which might encode the biases inherent to an LLM. To avoid this, we manually filter the prompt templates generated by LLM.

\input{noise_appendix}

\section{Prompt scores}
\label{sec:prompt_scores}
Table~\ref{tab:prompt_score_dataset} shows scores for the top 10 prompts selected during zero-shot evaluation of \method for each dataset (Note that all the scores are softmax normalized).
\onecolumn
\include{prompt_dataset_score}

\end{document}

%% file: algorithm.tex
\begin{algorithm}[t]
\caption{Weighted Prompt Ensemble}
\label{alg:wpe}
\begin{algorithmic}[]
\REQUIRE Downstream task $D$; Audio Representations $A$; Textual Representation, $T$ for $N$ prompts with $M$ class labels; Prompt scores $w \leftarrow \{w_1,\dots,w_N\}$; $N$ no of prompts: $p \in \{p\textsubscript{1},p\textsubscript{1},...,p\textsubscript{N}\}$
\FOR{$p \le N$}
    \STATE Compute prediction logits $\mathcal{L}_p \leftarrow A\cdot T_p^T$
    \STATE Compute max prediction logits: $\mathcal{L}^{max}_p \leftarrow max_j \, \mathcal{L}_p[i,j]$.
    \STATE Compute prompt score $w_p \leftarrow \sum_{i=1}^{|D|} \mathcal{L}^{max}_{p,d}$
    \STATE Update prompt score $w \leftarrow w \cup \{w_p\}$  
\ENDFOR
\STATE Normalize prompt scores $w \leftarrow \text{Softmax}(w)$
\STATE Perform weighted prompt ensemble: $T_{\text{avg}} \leftarrow \sum_{p=1}^N w_p \times T_p$ 
\end{algorithmic}
\end{algorithm}

%% file: main_results_table.tex
\begin{table*}[t]
\resizebox{1.0\textwidth}{!}{
\begin{tabular}{lcccccccccccc}
\toprule
\textbf{Model $\rightarrow$}  & \multicolumn{2}{c}{\textbf{L-CLAP}} & \multicolumn{2}{c}{\textbf{LM-CLAP}} & \multicolumn{2}{c}{\textbf{MSCLAP-22}} & \multicolumn{2}{c}{\textbf{MSCLAP-23}} & \multicolumn{2}{c}{\textbf{Wav2CLIP}} & \multicolumn{2}{c}{\textbf{CompA}} \\ \cline{2-13} 
\textbf{Dataset $\downarrow$} & ZS              & \method           & ZS               & \method           & ZS                & \method            & ZS                & \method            & ZS               & \method            & ZS              & \method          \\ \midrule
\multicolumn{13}{c}{Sound}\\ \midrule
ESC-50                        & 89.00           & \cellcolor{gray!10}\textbf{93.00}\textsubscript{\tiny{\textcolor{green}{+4.00\%}}}             & 85.60            & \cellcolor{gray!10}\textbf{92.65}\textsubscript{\tiny{\textcolor{green}{+7.05\%}}}             & 76.95             & \cellcolor{gray!10}\textbf{78.35}\textsubscript{\tiny{\textcolor{green}{+1.40\%}}}              & 91.80             & \cellcolor{gray!10}\textbf{94.80}\textsubscript{\tiny{\textcolor{green}{+3.00\%}}}              & 24.85            & \cellcolor{gray!10}\textbf{31.60}\textsubscript{\tiny{\textcolor{green}{+6.08\%}}}              & 91.35           & \cellcolor{gray!10}\textbf{93.20}\textsubscript{\tiny{\textcolor{green}{+1.85\%}}}            \\
USD-8K                        & 76.00           & \cellcolor{gray!10}\textbf{80.00}\textsubscript{\tiny{\textcolor{green}{+4.00\%}}}             & 28.09            & \cellcolor{gray!10}\textbf{39.93}\textsubscript{\tiny{\textcolor{green}{+11.84\%}}}             & 72.54             & \cellcolor{gray!10}\textbf{74.80}\textsubscript{\tiny{\textcolor{green}{+2.26\%}}}              & 77.70             & \cellcolor{gray!10}\textbf{82.50}\textsubscript{\tiny{\textcolor{green}{+4.80\%}}}              & 20.97            & \cellcolor{gray!10}\textbf{22.69}\textsubscript{\tiny{\textcolor{green}{+1.72\%}}}              & 73.53           & \cellcolor{gray!10}\textbf{78.32}\textsubscript{\tiny{\textcolor{green}{+4.79\%}}}            \\
TUT                           & 36.00           & \cellcolor{gray!10}\textbf{39.00}\textsubscript{\tiny{\textcolor{green}{+3.00\%}}}             & 28.09            & \cellcolor{gray!10}\textbf{39.93}\textsubscript{\tiny{\textcolor{green}{+11.84\%}}}             & 24.44             & \cellcolor{gray!10}\textbf{25.61}\textsubscript{\tiny{\textcolor{green}{+1.17\%}}}              & 45.00             & \cellcolor{gray!10}\textbf{47.00}\textsubscript{\tiny{\textcolor{green}{+2.00\%}}}              & 11.54            & \cellcolor{gray!10}\textbf{15.18}\textsubscript{\tiny{\textcolor{green}{+3.64\%}}}              & 40.12           & \cellcolor{gray!10}\textbf{46.28}\textsubscript{\tiny{\textcolor{green}{+6.16\%}}}            \\
VS                            & 78.20           & \cellcolor{gray!10}\textbf{80.00}\textsubscript{\tiny{\textcolor{green}{+1.80\%}}}             & 74.46            & \cellcolor{gray!10}78.91\textsubscript{\tiny{\textcolor{green}{+4.45\%}}}             & 43.78             & \cellcolor{gray!10}\textbf{54.94}\textsubscript{\tiny{\textcolor{green}{+11.16\%}}}              & 79.00             & \cellcolor{gray!10}\textbf{79.60}\textsubscript{\tiny{\textcolor{green}{+0.60\%}}}              & 22.72            & \cellcolor{gray!10}\textbf{24.06}\textsubscript{\tiny{\textcolor{green}{+1.31\%}}}              & 65.22           & \cellcolor{gray!10}\textbf{71.26}\textsubscript{\tiny{\textcolor{green}{+6.04\%}}}            \\
DCASE                         & 44.88           & \cellcolor{gray!10}\textbf{50.81}\textsubscript{\tiny{\textcolor{green}{+5.93\%}}}             & \textbf{56.76}            & \cellcolor{gray!10}55.94\textsubscript{\tiny{\textcolor{red}{-0.82\%}}}             & 13.93             & \cellcolor{gray!10}\textbf{23.77}\textsubscript{\tiny{\textcolor{green}{+9.84\%}}}              & 45.90             & \cellcolor{gray!10}\textbf{45.96}\textsubscript{\tiny{\textcolor{green}{+0.06\%}}}              & 09.63            & \cellcolor{gray!10}\textbf{17.21}\textsubscript{\tiny{\textcolor{green}{+7.58\%}}}              & 33.20           & \cellcolor{gray!10}\textbf{34.29}\textsubscript{\tiny{\textcolor{green}{+1.09\%}}}            \\
Gunshot Tri                   & 10.23           & \cellcolor{gray!10}\textbf{22.72}\textsubscript{\tiny{\textcolor{green}{+12.49\%}}}             & 13.64            & \cellcolor{gray!10}\textbf{29.52}\textsubscript{\tiny{\textcolor{green}{+15.88\%}}}             & 17.05             & \cellcolor{gray!10}\textbf{23.86}\textsubscript{\tiny{\textcolor{green}{+6.81\%}}}              & 25.00             & \cellcolor{gray!10}\textbf{25.00}\textsubscript{\tiny{\textcolor{green}{+0.00\%}}}              & \textbf{25.00}            & \cellcolor{gray!10}25.00\textsubscript{\tiny{\textcolor{red}{+0.00\%}}}              & 25.00           & \cellcolor{gray!10}\textbf{26.15}\textsubscript{\tiny{\textcolor{green}{+1.15\%}}}            \\
SESA                          & 67.72           & \cellcolor{gray!10}\textbf{74.28}\textsubscript{\tiny{\textcolor{green}{+6.56\%}}}             & 72.38            & \cellcolor{gray!10}\textbf{79.04}\textsubscript{\tiny{\textcolor{green}{+6.66\%}}}             & 66.67             & \cellcolor{gray!10}\textbf{68.47}\textsubscript{\tiny{\textcolor{green}{+1.80\%}}}              & 70.48             & \cellcolor{gray!10}\textbf{71.61}\textsubscript{\tiny{\textcolor{green}{+1.13\%}}}              & 29.52            & \cellcolor{gray!10}\textbf{56.10}\textsubscript{\tiny{\textcolor{green}{+26.58\%}}}              & 64.76           & \cellcolor{gray!10}\textbf{69.42}\textsubscript{\tiny{\textcolor{green}{+4.66\%}}}            \\
AudioSet                          & 31.88           & \cellcolor{gray!10}\textbf{36.98}\textsubscript{\tiny{\textcolor{green}{+5.10\%}}}             & 33.12            & \cellcolor{gray!10}38.21\textsubscript{\tiny{\textcolor{green}{+5.09\%}}}             & 16.10             & \cellcolor{gray!10}\textbf{17.81}\textsubscript{\tiny{\textcolor{green}{+1.71\%}}}              & 25.33             & \cellcolor{gray!10}\textbf{28.73}\textsubscript{\tiny{\textcolor{green}{+3.40\%}}}              & 18.03            & \cellcolor{gray!10}\textbf{20.12}\textsubscript{\tiny{\textcolor{green}{+2.09\%}}}              & 33.24           & \cellcolor{gray!10}\textbf{35.12}\textsubscript{\tiny{\textcolor{green}{+1.88\%}}}            \\
FSD50K                          & 46.45           & \cellcolor{gray!10}\textbf{48.76}\textsubscript{\tiny{\textcolor{green}{+2.31\%}}}             & 47.12            & \cellcolor{gray!10}\textbf{49.10}\textsubscript{\tiny{\textcolor{green}{+2.08\%}}}             & 32.50             & \cellcolor{gray!10}\textbf{33.80}\textsubscript{\tiny{\textcolor{green}{+1.30\%}}}              & 44.49             & \cellcolor{gray!10}\textbf{45.52}\textsubscript{\tiny{\textcolor{green}{+1.02\%}}}              & 42.31            & \cellcolor{gray!10}\textbf{44.14}\textsubscript{\tiny{\textcolor{green}{+2.07\%}}}              & 42.18           & \cellcolor{gray!10}\textbf{43.22}\textsubscript{\tiny{\textcolor{green}{+1.04\%}}}            \\
Cochlscene                    & 38.56           & \cellcolor{gray!10}\textbf{48.66}\textsubscript{\tiny{\textcolor{green}{+10.10\%}}}             & 50.66            & \cellcolor{gray!10}55.35\textsubscript{\tiny{\textcolor{green}{+4.69\%}}}             & 25.94             & \cellcolor{gray!10}\textbf{33.51}\textsubscript{\tiny{\textcolor{green}{+7.57\%}}}              & 85.00             & \cellcolor{gray!10}\textbf{85.22}\textsubscript{\tiny{\textcolor{green}{+0.22\%}}}              & 13.09            & \cellcolor{gray!10}\textbf{16.11}\textsubscript{\tiny{\textcolor{green}{+3.02\%}}}              & 31.95           & \cellcolor{gray!10}\textbf{38.21}\textsubscript{\tiny{\textcolor{green}{+6.26\%}}}            \\ \midrule \multicolumn{13}{c}{Music}\\ \midrule
Beijing Op.                   & 45.34           & \cellcolor{gray!10}\textbf{68.64}\textsubscript{\tiny{\textcolor{green}{+23.30\%}}}             & 75.00            & \cellcolor{gray!10}\textbf{75.42}\textsubscript{\tiny{\textcolor{green}{+0.42\%}}}             & 54.24             & \cellcolor{gray!10}\textbf{73.72}\textsubscript{\tiny{\textcolor{green}{+19.48\%}}}              & 71.19             & \cellcolor{gray!10}\textbf{71.61}\textsubscript{\tiny{\textcolor{green}{+0.42\%}}}              & 26.69            & \cellcolor{gray!10}\textbf{34.32}\textsubscript{\tiny{\textcolor{green}{+7.63\%}}}              & 61.86           & \cellcolor{gray!10}\textbf{63.21}\textsubscript{\tiny{\textcolor{green}{+1.35\%}}}            \\
GTZAN                         & 43.40           & \cellcolor{gray!10}\textbf{54.20}\textsubscript{\tiny{\textcolor{green}{+10.80\%}}}             & 63.92            & \cellcolor{gray!10}\textbf{63.93}\textsubscript{\tiny{\textcolor{green}{+0.01\%}}}             & 19.19             & \cellcolor{gray!10}\textbf{20.75}\textsubscript{\tiny{\textcolor{green}{+1.56\%}}}              & 56.24             & \cellcolor{gray!10}\textbf{58.56}\textsubscript{\tiny{\textcolor{green}{+2.32\%}}}              & \textbf{30.00}            & \cellcolor{gray!10}27.76\textsubscript{\tiny{\textcolor{red}{-2.24\%}}}              & 50.22           & \cellcolor{gray!10}\textbf{52.17}\textsubscript{\tiny{\textcolor{green}{+1.95\%}}}            \\
MUSDB                          & 55.60           & \cellcolor{gray!10}\textbf{66.00}\textsubscript{\tiny{\textcolor{green}{+10.40\%}}}             & 73.20            & \cellcolor{gray!10}\textbf{73.20}\textsubscript{\tiny{\textcolor{green}{+0.00\%}}}             & 47.20             & \cellcolor{gray!10}\textbf{47.75}\textsubscript{\tiny{\textcolor{green}{+0.55\%}}}              & 61.20             & \cellcolor{gray!10}\textbf{62.40}\textsubscript{\tiny{\textcolor{green}{+1.20\%}}}              & 51.60            & \cellcolor{gray!10}\textbf{52.20}\textsubscript{\tiny{\textcolor{green}{+0.60\%}}}              & 56.80           & \cellcolor{gray!10}\textbf{59.55}\textsubscript{\tiny{\textcolor{green}{+2.75\%}}}            \\
Medley                        & 82.50           & \cellcolor{gray!10}\textbf{92.00}\textsubscript{\tiny{\textcolor{green}{+9.50\%}}}             & 87.88            & \cellcolor{gray!10}94.30\textsubscript{\tiny{\textcolor{green}{+6.42\%}}}             & 84.41             & \cellcolor{gray!10}\textbf{86.20}\textsubscript{\tiny{\textcolor{green}{+1.79\%}}}              & 45.00             & \cellcolor{gray!10}\textbf{47.00}\textsubscript{\tiny{\textcolor{green}{+2.00\%}}}              & 42.20            & \cellcolor{gray!10}\textbf{47.08}\textsubscript{\tiny{\textcolor{green}{+4.88\%}}}              & 56.27           & \cellcolor{gray!10}\textbf{57.24}\textsubscript{\tiny{\textcolor{green}{+0.97\%}}}            \\
Mri. St                       & 10.81           & \cellcolor{gray!10}\textbf{37.35}\textsubscript{\tiny{\textcolor{green}{+26.54\%}}}             & 47.40            & \cellcolor{gray!10}\textbf{47.80}\textsubscript{\tiny{\textcolor{green}{+0.40\%}}}             & 14.50             & \cellcolor{gray!10}\textbf{14.80}\textsubscript{\tiny{\textcolor{green}{+0.30\%}}}              & 44.09             & \cellcolor{gray!10}\textbf{47.12}\textsubscript{\tiny{\textcolor{green}{+3.03\%}}}              & 06.09            & \cellcolor{gray!10}\textbf{19.49}\textsubscript{\tiny{\textcolor{green}{+13.40\%}}}              & 06.25           & \cellcolor{gray!10}\textbf{07.42}\textsubscript{\tiny{\textcolor{green}{+1.17\%}}}            \\
Mri. To                       & 25.10           & \cellcolor{gray!10}\textbf{34.38}\textsubscript{\tiny{\textcolor{green}{+9.28\%}}}             & 27.59            & \cellcolor{gray!10}\textbf{31.62}\textsubscript{\tiny{\textcolor{green}{+4.03\%}}}             & 16.50             & \cellcolor{gray!10}\textbf{16.63}\textsubscript{\tiny{\textcolor{green}{+0.13\%}}}              & 22.02             & \cellcolor{gray!10}\textbf{26.18}\textsubscript{\tiny{\textcolor{green}{+4.16\%}}}              & 15.57            & \cellcolor{gray!10}\textbf{24.95}\textsubscript{\tiny{\textcolor{green}{+9.38\%}}}              & 17.43           & \cellcolor{gray!10}\textbf{18.79}\textsubscript{\tiny{\textcolor{green}{+1.36\%}}}            \\
NSynth Inst                   & 37.20           & \cellcolor{gray!10}\textbf{38.00}\textsubscript{\tiny{\textcolor{green}{+0.80\%}}}             & 31.67            & \cellcolor{gray!10}\textbf{36.49}\textsubscript{\tiny{\textcolor{green}{+4.82\%}}}             & 26.26             & \cellcolor{gray!10}\textbf{29.63}\textsubscript{\tiny{\textcolor{green}{+3.37\%}}}              & 63.30             & \cellcolor{gray!10}\textbf{66.30}\textsubscript{\tiny{\textcolor{green}{+3.00\%}}}              & \textbf{24.39}            & \cellcolor{gray!10}21.72\textsubscript{\tiny{\textcolor{red}{-2.67\%}}}              & 27.86           & \cellcolor{gray!10}\textbf{29.24}\textsubscript{\tiny{\textcolor{green}{+1.38\%}}}            \\
NSynth Src                    & 37.00           & \cellcolor{gray!10}\textbf{41.00}\textsubscript{\tiny{\textcolor{green}{+4.00\%}}}             & 43.92            & \cellcolor{gray!10}\textbf{46.38}\textsubscript{\tiny{\textcolor{green}{+2.46\%}}}             & 37.06             & \cellcolor{gray!10}\textbf{41.45}\textsubscript{\tiny{\textcolor{green}{+4.39\%}}}              & 49.70             & \cellcolor{gray!10}\textbf{61.45}\textsubscript{\tiny{\textcolor{green}{+11.75\%}}}              & 38.28            & \cellcolor{gray!10}\textbf{42.01}\textsubscript{\tiny{\textcolor{green}{+3.73\%}}}              & 53.66           & \cellcolor{gray!10}\textbf{55.97}\textsubscript{\tiny{\textcolor{green}{+2.31\%}}}            \\ \bottomrule
\end{tabular}}
\caption{\small Performance comparison between \method and vanilla zero-shot classification (ZS) across 6 ALEs and 16 diverse audio classification tasks, including 10 sound and 8 musical datasets. The best scores for each ALE are in \textbf{bold}. Overall, \method outperforms vanilla ZS with improvements ranging from 0.42\% to 27\%.}
\label{tab:main_result}
\end{table*}

%% file: ablation_method_table.tex
\begin{table}[t]
\centering
\resizebox{1.0\columnwidth}{!}{
\begin{tabular}{l c}
\midrule
\textbf{Method}               & \textbf{Average Accuracy} \\ \midrule
\cellcolor{gray!10}MS-CLAP                       & \cellcolor{gray!10}58.28 \\ 
MS-CLAP+\small PE                    & 58.91 \\
MS-CLAP+\small WPE                   & 59.23 \\
MS-CLAP+\small PE+\small CMA                & 59.32 \\
MS-CLAP+ \method (\small WPE+\small CMA)      & \textbf{60.76} \\ \midrule
\end{tabular}
}
\caption{\small Performance comparison of MS-CLAP using various components of \method including Prompt Ensemble (PE), Weighted Prompt Ensemble (WPE), Cross-Modal Alignment (CMA) and their combinations. Baseline ZS scores are highlighted in grey. Overall, WPE consistently outperforms PE, both with and without CMA.}
\label{tab:ms_clap_results}
\vspace{-0.5em}
\end{table}

%% file: cma_analysis_table.tex
\begin{table}[t]
\centering
\resizebox{\columnwidth}{!}{
\begin{tabular}{ccccc}
\toprule
\textbf{Logits} & \textbf{Audio Guided} & \textbf{Text Guided} & \textbf{Average Accuracy} \\ \midrule
\cellcolor{gray!10}\textcolor{green}{$\checkmark$} &  \cellcolor{gray!10}\textcolor{red}{$\times$} & \cellcolor{gray!10}\textcolor{red}{$\times$} & \cellcolor{gray!10}58.28  \\
\textcolor{green}{$\checkmark$}  & \textcolor{red}{$\times$} & \textcolor{green}{$\checkmark$} & 59.24 \\
\textcolor{green}{$\checkmark$} & \textcolor{green}{$\checkmark$} & \textcolor{red}{$\times$} & 59.55 \\
\textcolor{red}{$\times$} & \textcolor{green}{$\checkmark$} & \textcolor{green}{$\checkmark$} & 58.29 \\
\textcolor{green}{$\checkmark$} & \textcolor{green}{$\checkmark$} & \textcolor{green}{$\checkmark$} & \textbf{60.76} \\ \bottomrule
\end{tabular}}
\caption{\small Performance comparison of MS-CLAP with \method across different combinations of original, audio-guided, and text-guided logits with Cross-Modal Alignment (CMA). Baseline ZS scores are highlighted in grey. Overall, incorporating all three logits in the final prediction yields the best zero-shot performance.}
\label{tab:cma}
\vspace{-0.5em}
\end{table}

%% file: prompt_number.tex
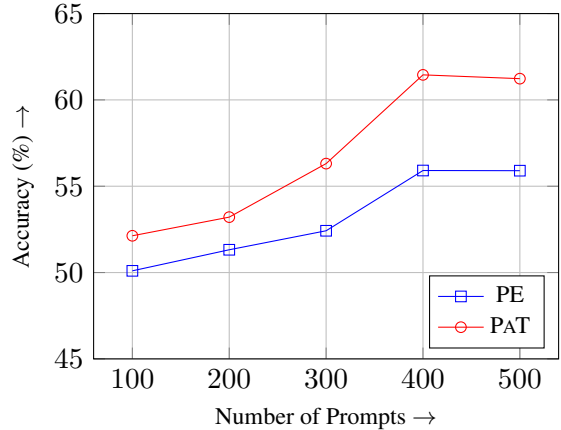
\begin{figure}
    \raggedright
    \begin{tikzpicture}
        \begin{axis}[
            xlabel={Number of Prompts $\rightarrow$},
            ylabel={Accuracy (\%) $\rightarrow$},
            legend pos=south east,
            grid=major,
            width=\columnwidth,
            height=0.8\columnwidth,
            ymin=45, ymax=65,
            xtick={100,200,300,400,500},
            ytick={45,50,55,60,65},
            legend style={font=\small},
            xlabel style={font=\small},
            ylabel style={font=\small},
        ]
        
        \addplot[
            color=blue,
            mark=square,
            ]
            coordinates {
            (100, 50.10) (200, 51.32) (300, 52.42) (400, 55.91) (500, 55.90)
            };
        \addlegendentry{PE}
        
        \addplot[
            color=red,
            mark=o,
            ]
            coordinates {
            (100, 52.13) (200, 53.21) (300, 56.31) (400, 61.45) (500, 61.23)
            };
        \addlegendentry{\method}

        \end{axis}
    \end{tikzpicture}
    \caption{\small Measuring MSCLAP-23 zero-shot performance with the prompt ensemble (PE) vs. \method on NSynth-Src by varying the prompt count. Generally, \method outperforms PE across different prompt counts.}
    \label{fig:wer_comparison}
\end{figure}

%% file: prompt_score.tex
\begin{table}[t]
\centering
\resizebox{1.0\columnwidth}{!}{
\begin{tabular}{ccc}
\toprule
\textbf{Dataset} & \textbf{Prompt} & \textbf{Score} \\
\midrule
ESC-50 & The sound of <label> coming from a cliff edge. & 0.0035 \\
 & A sound of a <label> coming from a parking lot & 0.0033 \\
\midrule
NSynth Inst & A major sound of a <label> & 0.0038 \\
& A minimal sound of a <label> & 0.0037 \\
\bottomrule
\end{tabular}}
\caption{\small Top two highest scoring prompt by \method for MSCLAP-23 on ESC-50 and Nsynth-Inst}
\label{tab:prompt_score}
\vspace{-0.5em}
\end{table}

%% file: noise_table.tex
\begin{table*}[t]
\resizebox{1.0\textwidth}{!}{
\begin{tabular}{@{}lcccccccccc@{}}
\toprule
\multirow{2}{*}{Dataset}            & \multicolumn{2}{c}{Gaussian Noise}         & \multicolumn{2}{c}{Pitch Shift} & \multicolumn{2}{c}{Polarity Inversion} & \multicolumn{2}{c}{Gain}        & \multicolumn{2}{c}{High Pass}   \\\cmidrule(l){2-3}\cmidrule(l){4-5}\cmidrule(l){6-7}\cmidrule(l){8-9}\cmidrule(l){10-11}
         & ZS        & \method            & ZS        & \method            & ZS            & \method               & ZS        & \method            & ZS        & \method            \\
\midrule \multicolumn{11}{c}{Sound} \\ \midrule
ESC-50       & 91.80          & \cellcolor{gray!10}\textbf{94.20}\textsubscript{\tiny{\textcolor{green}{+2.40\%}}} & 78.05          & \cellcolor{gray!10}\textbf{80.10}\textsubscript{\tiny{\textcolor{green}{+2.05\%}}} & 91.85              & \cellcolor{gray!10}\textbf{94.40}\textsubscript{\tiny{\textcolor{green}{+2.55\%}}}    & 92.05          & \cellcolor{gray!10}\textbf{94.85}\textsubscript{\tiny{\textcolor{green}{+2.80\%}}} & 82.35          & \cellcolor{gray!10}\textbf{86.15}\textsubscript{\tiny{\textcolor{green}{+3.80\%}}} \\
USD8K        & 77.26          & \cellcolor{gray!10}\textbf{82.70}\textsubscript{\tiny{\textcolor{green}{+5.44\%}}} & 63.61          & \cellcolor{gray!10}\textbf{70.31}\textsubscript{\tiny{\textcolor{green}{+6.70\%}}} & 77.43              & \cellcolor{gray!10}\textbf{82.69}\textsubscript{\tiny{\textcolor{green}{+5.26\%}}}    & 77.08          & \cellcolor{gray!10}\textbf{82.67}\textsubscript{\tiny{\textcolor{green}{+5.59\%}}} & 71.12          & \cellcolor{gray!10}\textbf{76.77}\textsubscript{\tiny{\textcolor{green}{+5.65\%}}} \\
TUT          & 44.94          & \cellcolor{gray!10}\textbf{45.74}\textsubscript{\tiny{\textcolor{green}{+0.80\%}}} & \textbf{26.05} & \cellcolor{gray!10}26.04\textsubscript{\tiny{\textcolor{red}{-0.01\%}}}          & 45.68              & \cellcolor{gray!10}\textbf{47.34}\textsubscript{\tiny{\textcolor{green}{+1.66\%}}}    & 38.95          & \cellcolor{gray!10}\textbf{41.97}\textsubscript{\tiny{\textcolor{green}{+3.02\%}}} & \textbf{35.80} & \cellcolor{gray!10}35.00\textsubscript{\tiny{\textcolor{red}{-0.80\%}}}          \\
VS           & \textbf{81.31} & \cellcolor{gray!10}77.86\textsubscript{\tiny{\textcolor{red}{-3.45\%}}}          & \textbf{76.61} & \cellcolor{gray!10}69.64\textsubscript{\tiny{\textcolor{red}{-6.97\%}}}          & \textbf{78.98}     & \cellcolor{gray!10}78.00\textsubscript{\tiny{\textcolor{red}{-0.98\%}}}             & 79.00          & \cellcolor{gray!10}\textbf{79.44}\textsubscript{\tiny{\textcolor{green}{+0.44\%}}} & 74.07          & \cellcolor{gray!10}\textbf{76.16}\textsubscript{\tiny{\textcolor{green}{+2.09\%}}} \\
DCASE        & 38.32          & \cellcolor{gray!10}\textbf{42.21}\textsubscript{\tiny{\textcolor{green}{+3.89\%}}} & 31.76          & \cellcolor{gray!10}\textbf{34.01}\textsubscript{\tiny{\textcolor{green}{+2.25\%}}} & 38.93              & \cellcolor{gray!10}\textbf{45.69}\textsubscript{\tiny{\textcolor{green}{+6.76\%}}}    & 43.24          & \cellcolor{gray!10}\textbf{45.28}\textsubscript{\tiny{\textcolor{green}{+2.04\%}}} & 33.40          & \cellcolor{gray!10}\textbf{37.70}\textsubscript{\tiny{\textcolor{green}{+4.30\%}}} \\
Gunshot Tri. & \textbf{25.00} & \cellcolor{gray!10}\textbf{25.00}\textsubscript{\tiny{\textcolor{green}{+0.00\%}}} & \textbf{25.00} & \cellcolor{gray!10}\textbf{25.00}\textsubscript{\tiny{\textcolor{green}{+0.00\%}}} & \textbf{25.00}     & \cellcolor{gray!10}\textbf{25.00}\textsubscript{\tiny{\textcolor{green}{+0.00\%}}}    & \textbf{25.00} & \cellcolor{gray!10}\textbf{25.00}\textsubscript{\tiny{\textcolor{green}{+0.00\%}}} & 19.32          & \cellcolor{gray!10}\textbf{22.72}\textsubscript{\tiny{\textcolor{green}{+3.40\%}}} \\
SESA         & 67.62          & \cellcolor{gray!10}\textbf{69.52}\textsubscript{\tiny{\textcolor{green}{+1.90\%}}} & 62.86          & \cellcolor{gray!10}\textbf{64.76}\textsubscript{\tiny{\textcolor{green}{+1.90\%}}} & 67.62              & \cellcolor{gray!10}\textbf{69.52}\textsubscript{\tiny{\textcolor{green}{+1.90\%}}}    & 68.57          & \cellcolor{gray!10}\textbf{69.52}\textsubscript{\tiny{\textcolor{green}{+0.95\%}}} & 48.57          & \cellcolor{gray!10}\textbf{58.10}\textsubscript{\tiny{\textcolor{green}{+9.53\%}}} \\
AudioSet     & 30.40   & \cellcolor{gray!10}\textbf{31.15}\textsubscript{\tiny{\textcolor{green}{+0.75\%}}}  & 22.37     & \cellcolor{gray!10}\textbf{23.06}\textsubscript{\tiny{\textcolor{green}{+0.69\%}}}  & \textbf{30.40} & \cellcolor{gray!10}29.22\textsubscript{\tiny{\textcolor{red}{-1.18\%}}}  & 28.78   & \cellcolor{gray!10}\textbf{30.23}\textsubscript{\tiny{\textcolor{green}{+1.45\%}}}      & 23.89      & \cellcolor{gray!10}\textbf{24.38}\textsubscript{\tiny{\textcolor{green}{+0.49\%}}} \\
FSD50K       & 44.54      & \cellcolor{gray!10}\textbf{45.74}\textsubscript{\tiny{\textcolor{green}{+1.20\%}}}     & 37.16      & \cellcolor{gray!10}\textbf{43.87}\textsubscript{\tiny{\textcolor{green}{+6.71\%}}}   & 44.39           & \cellcolor{gray!10}\textbf{44.96}\textsubscript{\tiny{\textcolor{green}{+0.57\%}}}   & \textbf{44.56}         & \cellcolor{gray!10}43.79\textsubscript{\tiny{\textcolor{red}{-0.77\%}}} & 37.94        & \cellcolor{gray!10}\textbf{43.73}\textsubscript{\tiny{\textcolor{green}{+5.79\%}}} \\
Cochlscene   & \textbf{85.07} & \cellcolor{gray!10}84.36\textsubscript{\tiny{\textcolor{red}{-0.71\%}}}          & 60.18          & \cellcolor{gray!10}\textbf{61.42}\textsubscript{\tiny{\textcolor{green}{+1.24\%}}} & 85.07              & \cellcolor{gray!10}\textbf{85.17}\textsubscript{\tiny{\textcolor{green}{+0.10\%}}}    & 81.97          & \cellcolor{gray!10}\textbf{82.09}\textsubscript{\tiny{\textcolor{green}{+0.12\%}}} & 73.34          & \cellcolor{gray!10}\textbf{75.15}\textsubscript{\tiny{\textcolor{green}{+1.81\%}}} \\
\midrule \multicolumn{11}{c}{Music} \\ \midrule
Beijing Op. & 70.34  & \cellcolor{gray!10}\textbf{70.62}\textsubscript{\tiny{\textcolor{green}{+0.28\%}}}  & 61.02  & \cellcolor{gray!10}\textbf{62.74}\textsubscript{\tiny{\textcolor{green}{+1.72\%}}}  & 71.19  & \cellcolor{gray!10}\textbf{71.61}\textsubscript{\tiny{\textcolor{green}{+0.42\%}}}  & 69.49  & \cellcolor{gray!10}\textbf{69.61}\textsubscript{\tiny{\textcolor{green}{+0.12\%}}}  & \textbf{65.68}  & \cellcolor{gray!10}64.86\textsubscript{\tiny{\textcolor{red}{-0.82\%}}}    \\
GTZAN  & 55.77          & \cellcolor{gray!10}\textbf{58.53}\textsubscript{\tiny{\textcolor{green}{+2.76\%}}} & 47.56          & \cellcolor{gray!10}\textbf{50.38}\textsubscript{\tiny{\textcolor{green}{+2.82\%}}} & 56.43              & \cellcolor{gray!10}\textbf{58.26}\textsubscript{\tiny{\textcolor{green}{+1.83\%}}}    & 55.20          & \cellcolor{gray!10}\textbf{57.79}\textsubscript{\tiny{\textcolor{green}{+2.59\%}}} & 47.36          & \cellcolor{gray!10}\textbf{50.54}\textsubscript{\tiny{\textcolor{green}{+3.18\%}}} \\
MUSDB         & \textbf{63.60} & \cellcolor{gray!10}53.60\textsubscript{\tiny{\textcolor{red}{-10.00\%}}}         & 58.00          & \cellcolor{gray!10}\textbf{61.60}\textsubscript{\tiny{\textcolor{green}{+3.60\%}}} & \textbf{68.00}     & \cellcolor{gray!10}56.80\textsubscript{\tiny{\textcolor{red}{-11.20\%}}}            & \textbf{68.00} & \cellcolor{gray!10}58.40\textsubscript{\tiny{\textcolor{red}{-9.60\%}}}          & 46.80          & \cellcolor{gray!10}\textbf{55.20}\textsubscript{\tiny{\textcolor{green}{+8.40\%}}} \\
Medley         & \textbf{96.61}  & \cellcolor{gray!10}95.96\textsubscript{\tiny{\textcolor{red}{-0.65\%}}}  & 92.09  & \cellcolor{gray!10}\textbf{92.46}\textsubscript{\tiny{\textcolor{green}{+0.37\%}}}  & 95.98  & \cellcolor{gray!10}\textbf{96.42}\textsubscript{\tiny{\textcolor{green}{+0.44\%}}}  & 95.97  & \cellcolor{gray!10}\textbf{96.53}\textsubscript{\tiny{\textcolor{green}{+0.56\%}}}  & \textbf{93.37}  & \cellcolor{gray!10}90.94\textsubscript{\tiny{\textcolor{red}{-2.43\%}}} \\
Mri. St.     & 42.63          & \cellcolor{gray!10}\textbf{48.93}\textsubscript{\tiny{\textcolor{green}{+6.30\%}}} & 33.15          & \cellcolor{gray!10}\textbf{44.94}\textsubscript{\tiny{\textcolor{green}{+11.79\%}}} & 44.09              & \cellcolor{gray!10}\textbf{47.12}\textsubscript{\tiny{\textcolor{green}{+3.03\%}}}    & 41.31          & \cellcolor{gray!10}\textbf{44.67}\textsubscript{\tiny{\textcolor{green}{+3.36\%}}} & 34.73          & \cellcolor{gray!10}\textbf{37.30}\textsubscript{\tiny{\textcolor{green}{+2.57\%}}} \\
Mri. To.     & 24.97          & \cellcolor{gray!10}\textbf{26.61}\textsubscript{\tiny{\textcolor{green}{+1.64\%}}} & 13.54          & \cellcolor{gray!10}\textbf{17.12}\textsubscript{\tiny{\textcolor{green}{+3.58\%}}} & 22.02              & \cellcolor{gray!10}\textbf{26.18}\textsubscript{\tiny{\textcolor{green}{+4.16\%}}}    & 19.08          & \cellcolor{gray!10}\textbf{26.78}\textsubscript{\tiny{\textcolor{green}{+7.70\%}}} & \textbf{17.24} & \cellcolor{gray!10}16.59\textsubscript{\tiny{\textcolor{red}{-0.65\%}}}          \\
NSynth Inst  & 52.86          & \cellcolor{gray!10}\textbf{53.85}\textsubscript{\tiny{\textcolor{green}{+0.99\%}}} & 60.11          & \cellcolor{gray!10}\textbf{64.77}\textsubscript{\tiny{\textcolor{green}{+4.66\%}}} & 63.89              & \cellcolor{gray!10}\textbf{66.33}\textsubscript{\tiny{\textcolor{green}{+2.44\%}}}    & 61.89          & \cellcolor{gray!10}\textbf{64.62}\textsubscript{\tiny{\textcolor{green}{+2.73\%}}} & 46.22          & \cellcolor{gray!10}\textbf{47.87}\textsubscript{\tiny{\textcolor{green}{+1.65\%}}} \\
NSynth Src   & 39.75          & \cellcolor{gray!10}\textbf{47.85}\textsubscript{\tiny{\textcolor{green}{+8.10\%}}} & 49.44          & \cellcolor{gray!10}\textbf{59.37}\textsubscript{\tiny{\textcolor{green}{+9.93\%}}} & 49.76              & \cellcolor{gray!10}\textbf{61.45}\textsubscript{\tiny{\textcolor{green}{+11.69\%}}}   & 49.49          & \cellcolor{gray!10}\textbf{60.64}\textsubscript{\tiny{\textcolor{green}{+11.15\%}}} & 47.39          & \cellcolor{gray!10}\textbf{56.46}\textsubscript{\tiny{\textcolor{green}{+9.07\%}}} \\ \bottomrule
\end{tabular}}
\caption{\small Zero-shot performance measure of MSCLAP-23 using \method across 16 audio classification tasks under noisy setting. Each audio sample is subjected to 5 different varieties of audio augmentations. \method outperforms vanilla zero-shot (ZS) classification scores by showing an absolute improvement of 0.10--11.15\%.}
\label{tab:noise table}
\vspace{-0.5em}
\end{table*}

%% file: noise_appendix.tex
\begin{table}[t]
\centering
\resizebox{\columnwidth}{!}{
\begin{tabular}{lcccccc}
\toprule
\multirow{2}{*}{Dataset}             & \multicolumn{2}{c}{Delay}       & \multicolumn{2}{c}{Low Pass}    & \multicolumn{2}{c}{Reverb}      \\ \cmidrule(l){2-3}\cmidrule(l){4-5}\cmidrule(l){6-7}
         & MS CLAP        & PAT            & MS CLAP        & PAT            & MS CLAP        & PAT            \\ 
\midrule \multicolumn{7}{c}{Sound} \\ \midrule
ESC-50       & 90.10          & \cellcolor{gray!10}\textbf{93.45}\textsubscript{\tiny{\textcolor{green}{+3.35\%}}} & 86.30          & \cellcolor{gray!10}\textbf{89.80}\textsubscript{\tiny{\textcolor{green}{+3.50\%}}} & 90.00          & \cellcolor{gray!10}\textbf{93.25}\textsubscript{\tiny{\textcolor{green}{+3.25\%}}} \\
USD8K        & \textbf{76.52} & \cellcolor{gray!10}75.95\textsubscript{\tiny{\textcolor{red}{-0.57\%}}}          & 72.79          & \cellcolor{gray!10}\textbf{78.45}\textsubscript{\tiny{\textcolor{green}{+5.66\%}}} & 76.54          & \cellcolor{gray!10}\textbf{81.88}\textsubscript{\tiny{\textcolor{green}{+5.34\%}}} \\
TUT          & \textbf{42.72} & \cellcolor{gray!10}42.22\textsubscript{\tiny{\textcolor{red}{-0.50\%}}}          & 39.07          & \cellcolor{gray!10}\textbf{43.64}\textsubscript{\tiny{\textcolor{green}{+4.57\%}}} & 42.90          & \cellcolor{gray!10}\textbf{43.64}\textsubscript{\tiny{\textcolor{green}{+0.74\%}}} \\
VS           & 78.39          & \cellcolor{gray!10}\textbf{78.58}\textsubscript{\tiny{\textcolor{green}{+0.19\%}}} & \textbf{77.47} & \cellcolor{gray!10}76.10\textsubscript{\tiny{\textcolor{red}{-1.37\%}}}          & \textbf{76.47} & \cellcolor{gray!10}74.82\textsubscript{\tiny{\textcolor{red}{-1.65\%}}}          \\
DCASE        & 38.11          & \cellcolor{gray!10}\textbf{43.44}\textsubscript{\tiny{\textcolor{green}{+5.33\%}}} & 41.19          & \cellcolor{gray!10}\textbf{41.80}\textsubscript{\tiny{\textcolor{green}{+0.61\%}}} & 38.32          & \cellcolor{gray!10}\textbf{44.67}\textsubscript{\tiny{\textcolor{green}{+6.35\%}}} \\
Gunshot Tri. & \textbf{25.00} & \cellcolor{gray!10}\textbf{25.00}\textsubscript{\tiny{\textcolor{green}{0.00\%}}}  & \textbf{25.00} & \cellcolor{gray!10}\textbf{25.00}\textsubscript{\tiny{\textcolor{green}{0.00\%}}}  & \textbf{30.68} & \cellcolor{gray!10}27.27\textsubscript{\tiny{\textcolor{red}{-3.41\%}}}          \\
SESA         & 68.57          & \cellcolor{gray!10}\textbf{69.52}\textsubscript{\tiny{\textcolor{green}{+0.95\%}}} & 68.57          & \cellcolor{gray!10}\textbf{69.52}\textsubscript{\tiny{\textcolor{green}{+0.95\%}}} & 68.57          & \cellcolor{gray!10}70.48\textsubscript{\tiny{\textcolor{green}{+1.91\%}}}          \\

AudioSet & 28.76 &   \cellcolor{gray!10}\textbf{29.19}\textsubscript{\tiny{\textcolor{green}{+0.43\%}}}   & 27.05     & \cellcolor{gray!10}\textbf{28.35}\textsubscript{\tiny{\textcolor{green}{+1.30\%}}}     &  28.92     &  \cellcolor{gray!10}\textbf{29.43}\textsubscript{\tiny{\textcolor{green}{+0.51\%}}}     \\
FSD50K & 41.84 &  \cellcolor{gray!10}\textbf{43.76}\textsubscript{\tiny{\textcolor{green}{+1.92\%}}}   &   41.10   &   \cellcolor{gray!10}\textbf{43.71}\textsubscript{\tiny{\textcolor{green}{+2.61\%}}}   & 43.21      &  \cellcolor{gray!10}\textbf{43.88}\textsubscript{\tiny{\textcolor{green}{+0.67\%}}}     \\
Cochlscene   & 83.28          & \cellcolor{gray!10}\textbf{84.43}\textsubscript{\tiny{\textcolor{green}{+1.15\%}}} & 59.82          & \cellcolor{gray!10}\textbf{60.32}\textsubscript{\tiny{\textcolor{green}{+0.50\%}}} & 80.72          & \cellcolor{gray!10}\textbf{81.44}\textsubscript{\tiny{\textcolor{green}{+0.72\%}}} \\ 
\midrule \multicolumn{7}{c}{Music} \\ \midrule
Beijing Op. & 68.54  & 	\cellcolor{gray!10}\textbf{70.21}\textsubscript{\tiny{\textcolor{green}{+1.67\%}}}  & 	64.83  & 	\cellcolor{gray!10}\textbf{65.44}\textsubscript{\tiny{\textcolor{green}{+0.61\%}}}  & 	71.61  & 	\cellcolor{gray!10}\textbf{72.45}\textsubscript{\tiny{\textcolor{green}{+0.84\%}}}    \\
GTZAN  & 54.41          & \cellcolor{gray!10}\textbf{57.49}\textsubscript{\tiny{\textcolor{green}{+3.08\%}}} & \textbf{56.40} & \cellcolor{gray!10}56.36\textsubscript{\tiny{\textcolor{red}{-0.04\%}}}          & 55.34          & \cellcolor{gray!10}\textbf{57.33}\textsubscript{\tiny{\textcolor{green}{+1.99\%}}} \\
MUSDB         & \textbf{71.20} & \cellcolor{gray!10}60.00\textsubscript{\tiny{\textcolor{red}{-11.20\%}}}         & \textbf{62.40} & \cellcolor{gray!10}\textbf{62.60}\textsubscript{\tiny{\textcolor{green}{+0.20\%}}}  & \textbf{69.20} & \cellcolor{gray!10}58.80\textsubscript{\tiny{\textcolor{red}{-10.40\%}}}         \\
Medley       & \textbf{96.09} & \cellcolor{gray!10}95.98\textsubscript{\tiny{\textcolor{red}{-0.11\%}}}          & \textbf{93.40} & \cellcolor{gray!10}92.57\textsubscript{\tiny{\textcolor{red}{-0.83\%}}}          & \textbf{96.27} & \cellcolor{gray!10}95.86\textsubscript{\tiny{\textcolor{red}{-0.41\%}}}          \\
Mri. St.     & 45.59          & \cellcolor{gray!10}\textbf{46.36}\textsubscript{\tiny{\textcolor{green}{+0.77\%}}} & 37.54          & \cellcolor{gray!10}\textbf{44.16}\textsubscript{\tiny{\textcolor{green}{+6.62\%}}} & \textbf{33.51} & \cellcolor{gray!10}28.50\textsubscript{\tiny{\textcolor{red}{-5.01\%}}}          \\
Mri. To.     & 22.36          & \cellcolor{gray!10}\textbf{31.72}\textsubscript{\tiny{\textcolor{green}{+9.36\%}}} & 19.56          & \cellcolor{gray!10}\textbf{22.83}\textsubscript{\tiny{\textcolor{green}{+3.27\%}}} & 13.49          & \cellcolor{gray!10}\textbf{19.32}\textsubscript{\tiny{\textcolor{green}{+5.83\%}}} \\
NSynth Inst  & 59.94          & \cellcolor{gray!10}\textbf{61.66}\textsubscript{\tiny{\textcolor{green}{+1.72\%}}} & 53.91          & \cellcolor{gray!10}\textbf{57.00}\textsubscript{\tiny{\textcolor{green}{+3.09\%}}} & 56.76          & \cellcolor{gray!10}\textbf{59.27}\textsubscript{\tiny{\textcolor{green}{+2.51\%}}} \\
NSynth Src   & 48.12          & \cellcolor{gray!10}\textbf{58.10}\textsubscript{\tiny{\textcolor{green}{+9.98\%}}} & 39.06          & \cellcolor{gray!10}\textbf{49.75}\textsubscript{\tiny{\textcolor{green}{+10.69\%}}} & 47.24          & \cellcolor{gray!10}\textbf{55.73}\textsubscript{\tiny{\textcolor{green}{+8.49\%}}} \\ \bottomrule
\end{tabular}}
\caption{\small Zero-shot performance measure of MSCLAP-23 using \method across 16 audio classification tasks under noisy settings. Each audio sample is subjected to 3 different varieties of audio augmentations—delay, low pass filtering, and reverb. \method outperforms vanilla zero-shot (ZS) classification scores by showing an absolute improvement of 0.10--10.69\%.}
\label{tab:noise_appendix}
\end{table}

%% file: prompt_dataset_score.tex
\begin{longtable}{cll}

\midrule
Dataset & Prompt & Score 
\\
\toprule
\multirow{11}{*}{Beijing Op.} & A sound of a \textless sound\textgreater coming from a playground.                  & 0.00372 \\
& A sound of a \textless sound\textgreater coming from a parade.                      & 0.00365 \\
& A sound of a \textless sound\textgreater coming from a swimming pool.               & 0.00364 \\
& A sound of a \textless sound\textgreater coming from a park.                        & 0.00363 \\
& A restrained sound of a \textless sound\textgreater.  & 0.00362 \\
& A soft sound of a \textless sound\textgreater.    & 0.00362 \\
& The sound of \textless sound\textgreater can be heard near a playground.            & 0.00361 \\
& A sound of a \textless sound\textgreater coming from a zoo exhibit.                 & 0.0036  \\
& A sound of a \textless sound\textgreater coming from a gym.& 0.0036  \\
& A subtle sound of a \textless sound\textgreater.  & 0.0036  \\
& A gentle sound of a \textless sound\textgreater.  & 0.00359 \\ \midrule
\multirow{11}{*}{Cochlsene}                       & An even sound of a \textless sound\textgreater.   & 0.00357 \\
& The sound of \textless sound\textgreater can be heard near a rooftop garden.        & 0.00357 \\
& A low-key sound of a \textless sound\textgreater. & 0.00355 \\
& A sporadic sound of a \textless sound\textgreater.& 0.00355 \\
& An irregular sound of a \textless sound\textgreater.  & 0.00354 \\
& A subdued sound of a \textless sound\textgreater. & 0.00354 \\
& A delicate sound of a \textless sound\textgreater.& 0.00354 \\
& The sound of \textless sound\textgreater can be heard near a lighthouse.            & 0.00353 \\
& A moderate sound of a \textless sound\textgreater.& 0.00353 \\
& A quiet sound of a \textless sound\textgreater.   & 0.00353 \\
& A gentle sound of a \textless sound\textgreater.  & 0.00353 \\ \midrule
\multirow{11}{*}{DCASE}  & A reverberating sound of a \textless sound\textgreater.    & 0.00352 \\
& An extensive sound of a \textless sound\textgreater.  & 0.00352 \\
& A tremendous sound of a \textless sound\textgreater.  & 0.00352 \\
& An overwhelming sound of a \textless sound\textgreater.    & 0.00351 \\
& A sound of a \textless sound\textgreater coming from a church.                      & 0.0035  \\
& The sound of \textless sound\textgreater can be heard near a hedge maze.            & 0.00348 \\
& The sound of \textless sound\textgreater can be heard near a cliff edge.            & 0.00347 \\
& An enormous sound of a \textless sound\textgreater.   & 0.00347 \\
& A massive sound of a \textless sound\textgreater. & 0.00347 \\
& The sound of \textless sound\textgreater can be heard near a fishing pier.          & 0.00346 \\
& A vibrant sound of a \textless sound\textgreater. & 0.00346 \\ \midrule
\multirow{11}{*}{ESC50}  & An extensive sound of a \textless sound\textgreater.  & 0.00363 \\
& An unwavering sound of a \textless sound\textgreater. & 0.0036  \\
& A sporadic sound of a \textless sound\textgreater.& 0.0036  \\
& A persistent sound of a \textless sound\textgreater.  & 0.0036  \\
& An all-encompassing sound of a \textless sound\textgreater.& 0.00359 \\
& A sound of a \textless sound\textgreater coming from a garden.                      & 0.00359 \\
& A reverberating sound of a \textless sound\textgreater.    & 0.00358 \\
& A sound of a \textless sound\textgreater coming from a barber shop.                 & 0.00358 \\
& An even sound of a \textless sound\textgreater.   & 0.00358 \\
& A continuous sound of a \textless sound\textgreater.  & 0.00356 \\
& An extreme sound of a \textless sound\textgreater.& 0.00356 \\ \midrule
\multirow{6}{*}{GTZAN}  & A sound of a \textless sound\textgreater coming from a church.                      & 0.00379 \\
& The sound of \textless sound\textgreater can be heard near a church.                & 0.00374 \\
& A sound of a \textless sound\textgreater coming from a park.                        & 0.00374 \\
& The sound of \textless sound\textgreater can be heard near a concert hall.          & 0.00373 \\
& A sound of a \textless sound\textgreater coming from a theater.                     & 0.0037  \\
& A sound of a \textless sound\textgreater coming from a zoo.& 0.0037  \\
& The sound of \textless sound\textgreater can be heard near a fruit orchard.         & 0.00369 \\
& The sound of \textless sound\textgreater can be heard near a theater.               & 0.00369 \\
& A sound of a \textless sound\textgreater coming from a concert hall.                & 0.00368 \\
& An extensive sound of a \textless sound\textgreater.  & 0.00367 \\ 
& The sound of \textless sound\textgreater can be heard near a garden.                & 0.00367 \\ \midrule
\multirow{11}{*}{Gunshot Tri.}                    & The sound of \textless sound\textgreater can be heard near a forest trail.          & 0.00418 \\
& The sound of \textless sound\textgreater can be heard near a suburban neighborhood. & 0.00417 \\
& The sound of \textless sound\textgreater can be heard near a wildlife reserve.      & 0.00416 \\
& A sound of a \textless sound\textgreater coming from a sports field.                & 0.00409 \\
& A sound of a \textless sound\textgreater coming from a park.                        & 0.00407 \\
& The sound of \textless sound\textgreater can be heard near a golf course.           & 0.00407 \\
& The sound of \textless sound\textgreater can be heard near a lake.                  & 0.00406 \\
& A sound of a \textless sound\textgreater coming from a parking lot.                 & 0.00405 \\
& A sound of a \textless sound\textgreater coming from a forest.                      & 0.00401 \\
& The sound of \textless sound\textgreater can be heard near a greenfield.            & 0.00398 \\
& The sound of \textless sound\textgreater can be heard near a bridge.                & 0.00396 \\ \midrule
\multirow{11}{*}{Medley} & A gentle sound of a \textless sound\textgreater.  & 0.00393 \\
& A minor sound of a \textless sound\textgreater.   & 0.00389 \\
& A mild sound of a \textless sound\textgreater.    & 0.00387 \\
& A soft sound of a \textless sound\textgreater.    & 0.00385 \\
& A restrained sound of a \textless sound\textgreater.  & 0.00382 \\
& A feeble sound of a \textless sound\textgreater.  & 0.0038  \\
& A subdued sound of a \textless sound\textgreater. & 0.00379 \\
& An even sound of a \textless sound\textgreater.   & 0.00378 \\
& A delicate sound of a \textless sound\textgreater.& 0.00378 \\
& A major sound of a \textless sound\textgreater.   & 0.00376 \\
& A faint sound of a \textless sound\textgreater.   & 0.00375 \\ \midrule
\multirow{11}{*}{Mridangam St.}                   & A minimal sound of a \textless sound\textgreater. & 0.00413 \\
& A firm sound of a \textless sound\textgreater.    & 0.00408 \\
& A resounding sound of a \textless sound\textgreater.  & 0.00407 \\
& A muted sound of a \textless sound\textgreater.   & 0.00404 \\
& A robust sound of a \textless sound\textgreater.  & 0.00401 \\
& An even sound of a \textless sound\textgreater.   & 0.00399 \\
& A soft sound of a \textless sound\textgreater.    & 0.00395 \\
& A moderate sound of a \textless sound\textgreater.& 0.00393 \\
& A feeble sound of a \textless sound\textgreater.  & 0.00392 \\
& A major sound of a \textless sound\textgreater.   & 0.00389 \\
& A gentle sound of a \textless sound\textgreater.  & 0.00384 \\ \midrule
\multirow{11}{*}{Mridangam Tonic}                 & A minimal sound of a \textless sound\textgreater. & 0.004   \\
& A sound of a \textless sound\textgreater coming from a music box.                   & 0.00387 \\
& A sound of a \textless sound\textgreater coming from a car wash.                    & 0.00384 \\
& A sound of a \textless sound\textgreater coming from a dock.                        & 0.00383 \\
& A sound of a \textless sound\textgreater coming from a river.                       & 0.0038  \\
& A sound of a \textless sound\textgreater coming from a microwave.                   & 0.00378 \\
& The sound of \textless sound\textgreater can be heard near a vineyard.              & 0.00376 \\
& A sound of a \textless sound\textgreater coming from a clock.                       & 0.00375 \\
& A sound of a \textless sound\textgreater coming from a washing machine.             & 0.00375 \\
& A sound of a \textless sound\textgreater coming from a car engine.                  & 0.00371 \\
& A sound of a \textless sound\textgreater coming from a toy.& 0.00369 \\ \midrule
\multirow{11}{*}{MUSDB}  & A vibrant sound of a \textless sound\textgreater. & 0.00397 \\
& An even sound of a \textless sound\textgreater.   & 0.0039  \\
& A stentorian sound of a \textless sound\textgreater.  & 0.00389 \\
& A minor sound of a \textless sound\textgreater.   & 0.00383 \\
& A serene sound of a \textless sound\textgreater.  & 0.00379 \\
& A gentle sound of a \textless sound\textgreater.  & 0.00379 \\
& A sound of a \textless sound\textgreater coming from a theater.                     & 0.00377 \\
& A restrained sound of a \textless sound\textgreater.  & 0.00374 \\
& A quiet sound of a \textless sound\textgreater.   & 0.00374 \\
& An all-encompassing sound of a \textless sound\textgreater.& 0.00374 \\
& A major sound of a \textless sound\textgreater.   & 0.00373 \\ \midrule
\multirow{11}{*}{Nsynth Inst}                     & A major sound of a \textless sound\textgreater.   & 0.00389 \\
& A minimal sound of a \textless sound\textgreater. & 0.00387 \\
& A resonant sound of a \textless sound\textgreater.& 0.00385 \\
& A mild sound of a \textless sound\textgreater.    & 0.00384 \\
& A gentle sound of a \textless sound\textgreater.  & 0.0038  \\
& A minor sound of a \textless sound\textgreater.   & 0.00376 \\
& A moderate sound of a \textless sound\textgreater.& 0.00375 \\
& A sharp sound of a \textless sound\textgreater.   & 0.00373 \\
& A slight sound of a \textless sound\textgreater.  & 0.00372 \\
& A soft sound of a \textless sound\textgreater.    & 0.00372 \\
& A resounding sound of a \textless sound\textgreater.  & 0.00371 \\ \midrule
\multirow{11}{*}{Nsynth Source}                   & A resonant sound of a \textless sound\textgreater.& 0.00387 \\
& A robust sound of a \textless sound\textgreater.  & 0.00387 \\
& A minor sound of a \textless sound\textgreater.   & 0.00382 \\
& A moderate sound of a \textless sound\textgreater.& 0.00382 \\
& A sound of a \textless sound\textgreater coming from a piano.                       & 0.00382 \\
& A resounding sound of a \textless sound\textgreater.  & 0.0038  \\
& A firm sound of a \textless sound\textgreater.    & 0.0038  \\
& A mild sound of a \textless sound\textgreater.    & 0.00374 \\
& A slight sound of a \textless sound\textgreater.  & 0.00373 \\
& A sound of a \textless sound\textgreater coming from a guitar.                      & 0.00372 \\
& An even sound of a \textless sound\textgreater.   & 0.00371 \\ \midrule
\multirow{11}{*}{SESA}   & The sound of \textless sound\textgreater can be heard near a garden.                & 0.00364 \\
& A sound of a \textless sound\textgreater coming from a parking lot.                 & 0.00361 \\
& The sound of \textless sound\textgreater can be heard near a fishing pier.          & 0.00361 \\
& The sound of \textless sound\textgreater can be heard near a hedge maze.            & 0.0036  \\
& The sound of \textless sound\textgreater can be heard near a rooftop garden.        & 0.00358 \\
& A sound of a \textless sound\textgreater coming from a garden.                      & 0.00358 \\
& The sound of \textless sound\textgreater can be heard near a golf course.           & 0.00358 \\
& The sound of \textless sound\textgreater can be heard near a playground.            & 0.00357 \\
& The sound of \textless sound\textgreater can be heard near a university.            & 0.00357 \\
& A sound of a \textless sound\textgreater coming from a park.                        & 0.00357 \\
& An echoing sound of a \textless sound\textgreater.& 0.00357 \\ \midrule
\multirow{6}{*}{TUT}    & A quiet sound of a \textless sound\textgreater.   & 0.00372 \\
& A subdued sound of a \textless sound\textgreater. & 0.00367 \\
& A low-key sound of a \textless sound\textgreater. & 0.00366 \\
& A hushed sound of a \textless sound\textgreater.  & 0.00366 \\
& A faint sound of a \textless sound\textgreater.   & 0.00365 \\
& The sound of \textless sound\textgreater can be heard near a beach.                 & 0.00363 \\
& An even sound of a \textless sound\textgreater.   & 0.0036  \\
& A calm sound of a \textless sound\textgreater.    & 0.0036  \\
& A sound of a \textless sound\textgreater coming from a hallway.                     & 0.0036  \\
& A soft sound of a \textless sound\textgreater.    & 0.00359 \\
& A sound of a \textless sound\textgreater coming from a hospital room.               & 0.00359 \\ \midrule
\multirow{11}{*}{USD8K}  & A sound of a \textless sound\textgreater coming from a park.                        & 0.0036  \\
& A subtle sound of a \textless sound\textgreater.  & 0.0036  \\
& A soft sound of a \textless sound\textgreater.    & 0.0036  \\
& A mild sound of a \textless sound\textgreater.    & 0.0036  \\
& A slight sound of a \textless sound\textgreater.  & 0.00359 \\
& A feeble sound of a \textless sound\textgreater.  & 0.00358 \\
& A faint sound of a \textless sound\textgreater.   & 0.00357 \\
& A muted sound of a \textless sound\textgreater.   & 0.00357 \\
& The sound of \textless sound\textgreater can be heard near a university.            & 0.00357 \\
& A minimal sound of a \textless sound\textgreater. & 0.00357 \\
& A persistent sound of a \textless sound\textgreater.  & 0.00357 \\ \midrule
\multirow{11}{*}{Vocal Sound}                     & A sudden sound of a \textless sound\textgreater.  & 0.00377 \\
& A sound of a \textless sound\textgreater coming from a barber shop.                 & 0.00377 \\
& A quiet sound of a \textless sound\textgreater.   & 0.00377 \\
& An even sound of a \textless sound\textgreater.   & 0.00376 \\
& An abrupt sound of a \textless sound\textgreater. & 0.00375 \\
& A gentle sound of a \textless sound\textgreater.  & 0.00372 \\
& A low-key sound of a \textless sound\textgreater. & 0.00372 \\
& A faint sound of a \textless sound\textgreater.   & 0.00371 \\
& A sporadic sound of a \textless sound\textgreater.& 0.00371 \\
& A major sound of a \textless sound\textgreater.   & 0.0037  \\
& A subtle sound of a \textless sound\textgreater.  & 0.0037  \\ \bottomrule
\caption{Score of Top 10 prompts across various audio classification datasets}
\label{tab:prompt_score_dataset}
\end{longtable}

%% file: acl_latex.bbl
\begin{thebibliography}{41}
\providecommand{\natexlab}[1]{#1}

\bibitem[{Anantapadmanabhan et~al.(2014)Anantapadmanabhan, Bellur, and Murthy}]{anantapadmanabhan2014mridangam}
Akshay Anantapadmanabhan, Ashwin Bellur, and Hema~A. Murthy. 2014.
\newblock \href {https://doi.org/10.5281/zenodo.1265188} {Mridangam stroke dataset (1.0)}.
\newblock 2013 IEEE International Conference on Acoustics, Speech and Signal Processing, Vancouver, BC, Canada. Data set.

\bibitem[{Bertin-Mahieux et~al.(2011)Bertin-Mahieux, Ellis, Whitman, and Lamere}]{Bertin-Mahieux2011}
Thierry Bertin-Mahieux, Daniel~P.W. Ellis, Brian Whitman, and Paul Lamere. 2011.
\newblock The million song dataset.
\newblock In \emph{{Proceedings of the 12th International Conference on Music Information Retrieval ({ISMIR} 2011)}}.

\bibitem[{Bittner et~al.(2014)Bittner, Salamon, Tierney, Mauch, Cannam, and Bello}]{bittner2014medleydb}
Rachel~M Bittner, Justin Salamon, Mike Tierney, Matthias Mauch, Chris Cannam, and Juan~Pablo Bello. 2014.
\newblock Medleydb: A multitrack dataset for annotation-intensive mir research.
\newblock In \emph{ISMIR}, volume~14, pages 155--160.

\bibitem[{Brown(2020)}]{brown2020language}
Tom~B Brown. 2020.
\newblock Language models are few-shot learners.
\newblock \emph{arXiv preprint arXiv:2005.14165}.

\bibitem[{Elizalde et~al.(2023{\natexlab{a}})Elizalde, Deshmukh, Al~Ismail, and Wang}]{elizalde2023clap}
Benjamin Elizalde, Soham Deshmukh, Mahmoud Al~Ismail, and Huaming Wang. 2023{\natexlab{a}}.
\newblock Clap learning audio concepts from natural language supervision.
\newblock In \emph{ICASSP 2023-2023 IEEE International Conference on Acoustics, Speech and Signal Processing (ICASSP)}, pages 1--5. IEEE.

\bibitem[{Elizalde et~al.(2023{\natexlab{b}})Elizalde, Deshmukh, Al~Ismail, and Wang}]{CLAP2022}
Benjamin Elizalde, Soham Deshmukh, Mahmoud Al~Ismail, and Huaming Wang. 2023{\natexlab{b}}.
\newblock Clap learning audio concepts from natural language supervision.
\newblock In \emph{ICASSP 2023-2023 IEEE International Conference on Acoustics, Speech and Signal Processing (ICASSP)}, pages 1--5. IEEE.

\bibitem[{Engel et~al.(2017{\natexlab{a}})Engel, Resnick, Roberts, Dieleman, Eck, Simonyan, and Norouzi}]{nsynth2017}
Jesse Engel, Cinjon Resnick, Adam Roberts, Sander Dieleman, Douglas Eck, Karen Simonyan, and Mohammad Norouzi. 2017{\natexlab{a}}.
\newblock \href {https://arxiv.org/abs/arXiv:1704.01279} {Neural audio synthesis of musical notes with wavenet autoencoders}.

\bibitem[{Engel et~al.(2017{\natexlab{b}})Engel, Resnick, Roberts, Dieleman, Norouzi, Eck, and Simonyan}]{engel2017neural}
Jesse Engel, Cinjon Resnick, Adam Roberts, Sander Dieleman, Mohammad Norouzi, Douglas Eck, and Karen Simonyan. 2017{\natexlab{b}}.
\newblock Neural audio synthesis of musical notes with wavenet autoencoders.
\newblock In \emph{International Conference on Machine Learning}, pages 1068--1077. PMLR.

\bibitem[{Fonseca et~al.(2021)Fonseca, Favory, Pons, Font, and Serra}]{fonseca2021fsd50k}
Eduardo Fonseca, Xavier Favory, Jordi Pons, Frederic Font, and Xavier Serra. 2021.
\newblock Fsd50k: an open dataset of human-labeled sound events.
\newblock \emph{IEEE/ACM Transactions on Audio, Speech, and Language Processing}, 30:829--852.

\bibitem[{Gemmeke et~al.(2017)Gemmeke, Ellis, Freedman, Jansen, Lawrence, Moore, Plakal, and Ritter}]{gemmeke2017audio}
Jort~F Gemmeke, Daniel~PW Ellis, Dylan Freedman, Aren Jansen, Wade Lawrence, R~Channing Moore, Manoj Plakal, and Marvin Ritter. 2017.
\newblock Audio set: An ontology and human-labeled dataset for audio events.
\newblock In \emph{2017 IEEE international conference on acoustics, speech and signal processing (ICASSP)}, pages 776--780. IEEE.

\bibitem[{Ghosh et~al.(2024{\natexlab{a}})Ghosh, Kumar, Evuru, Nieto, Duraiswami, and Manocha}]{ghosh2024reclap}
Sreyan Ghosh, Sonal Kumar, Chandra Kiran~Reddy Evuru, Oriol Nieto, Ramani Duraiswami, and Dinesh Manocha. 2024{\natexlab{a}}.
\newblock Reclap: Improving zero shot audio classification by describing sounds.
\newblock \emph{arXiv preprint arXiv:2409.09213}.

\bibitem[{Ghosh et~al.(2024{\natexlab{b}})Ghosh, Kumar, Seth, Evuru, Tyagi, Sakshi, Nieto, Duraiswami, and Manocha}]{ghosh2024gama}
Sreyan Ghosh, Sonal Kumar, Ashish Seth, Chandra Kiran~Reddy Evuru, Utkarsh Tyagi, S~Sakshi, Oriol Nieto, Ramani Duraiswami, and Dinesh Manocha. 2024{\natexlab{b}}.
\newblock Gama: A large audio-language model with advanced audio understanding and complex reasoning abilities.
\newblock \emph{arXiv preprint arXiv:2406.11768}.

\bibitem[{Ghosh et~al.(2023)Ghosh, Seth, Kumar, Tyagi, Evuru, Ramaneswaran, Sakshi, Nieto, Duraiswami, and Manocha}]{ghosh2023compa}
Sreyan Ghosh, Ashish Seth, Sonal Kumar, Utkarsh Tyagi, Chandra~Kiran Evuru, S~Ramaneswaran, S~Sakshi, Oriol Nieto, Ramani Duraiswami, and Dinesh Manocha. 2023.
\newblock Compa: Addressing the gap in compositional reasoning in audio-language models.
\newblock \emph{arXiv preprint arXiv:2310.08753}.

\bibitem[{Gong et~al.(2022)Gong, Yu, and Glass}]{gong_vocalsound}
Yuan Gong, Jin Yu, and James Glass. 2022.
\newblock \href {https://doi.org/10.1109/ICASSP43922.2022.9746828} {Vocalsound: A dataset for improving human vocal sounds recognition}.
\newblock In \emph{ICASSP 2022 - 2022 IEEE International Conference on Acoustics, Speech and Signal Processing (ICASSP)}, pages 151--155.

\bibitem[{Guzhov et~al.(2022)Guzhov, Raue, Hees, and Dengel}]{guzhov2022audioclip}
Andrey Guzhov, Federico Raue, J{\"o}rn Hees, and Andreas Dengel. 2022.
\newblock Audioclip: Extending clip to image, text and audio.
\newblock In \emph{ICASSP 2022-2022 IEEE International Conference on Acoustics, Speech and Signal Processing (ICASSP)}, pages 976--980. IEEE.

\bibitem[{Hanif et~al.(2024)Hanif, Agro, Qazi, and Aldarmaki}]{hanif2024palm}
Asif Hanif, Maha~Tufail Agro, Mohammad~Areeb Qazi, and Hanan Aldarmaki. 2024.
\newblock Palm: Few-shot prompt learning for audio language models.
\newblock \emph{arXiv preprint arXiv:2409.19806}.

\bibitem[{Houlsby et~al.(2019)Houlsby, Giurgiu, Jastrzebski, Morrone, De~Laroussilhe, Gesmundo, Attariyan, and Gelly}]{houlsby2019parameter}
Neil Houlsby, Andrei Giurgiu, Stanislaw Jastrzebski, Bruna Morrone, Quentin De~Laroussilhe, Andrea Gesmundo, Mona Attariyan, and Sylvain Gelly. 2019.
\newblock Parameter-efficient transfer learning for nlp.
\newblock In \emph{International conference on machine learning}, pages 2790--2799. PMLR.

\bibitem[{Jeong and Park(2022)}]{jeong2022cochlscene}
Il-Young Jeong and Jeongsoo Park. 2022.
\newblock Cochlscene: Acquisition of acoustic scene data using crowdsourcing.
\newblock In \emph{2022 Asia-Pacific Signal and Information Processing Association Annual Summit and Conference (APSIPA ASC)}, pages 17--21. IEEE.

\bibitem[{Kim et~al.(2024)Kim, Jung, Lee, and Woo}]{kim2024enclap}
Jaeyeon Kim, Jaeyoon Jung, Jinjoo Lee, and Sang~Hoon Woo. 2024.
\newblock Enclap: Combining neural audio codec and audio-text joint embedding for automated audio captioning.
\newblock In \emph{ICASSP 2024-2024 IEEE International Conference on Acoustics, Speech and Signal Processing (ICASSP)}, pages 6735--6739. IEEE.

\bibitem[{Li et~al.(2024)Li, Wang, and Liu}]{10446472}
Yiming Li, Xiangdong Wang, and Hong Liu. 2024.
\newblock \href {https://doi.org/10.1109/ICASSP48485.2024.10446472} {Audio-free prompt tuning for language-audio models}.
\newblock In \emph{ICASSP 2024 - 2024 IEEE International Conference on Acoustics, Speech and Signal Processing (ICASSP)}, pages 491--495.

\bibitem[{Liang et~al.(2023)Liang, Liu, Liu, Phan, Benetos, Plumbley, and Wang}]{liang2023adapting}
Jinhua Liang, Xubo Liu, Haohe Liu, Huy Phan, Emmanouil Benetos, Mark~D Plumbley, and Wenwu Wang. 2023.
\newblock Adapting language-audio models as few-shot audio learners.
\newblock \emph{arXiv preprint arXiv:2305.17719}.

\bibitem[{Mesaros et~al.(2017{\natexlab{a}})Mesaros, Heittola, Diment, Elizalde, Shah, Vincent, Raj, and Virtanen}]{mesaros2017dcase}
Annamaria Mesaros, Toni Heittola, Aleksandr Diment, Benjamin Elizalde, Ankit Shah, Emmanuel Vincent, Bhiksha Raj, and Tuomas Virtanen. 2017{\natexlab{a}}.
\newblock Dcase 2017 challenge setup: Tasks, datasets and baseline system.
\newblock In \emph{DCASE 2017-Workshop on Detection and Classification of Acoustic Scenes and Events}.

\bibitem[{Mesaros et~al.(2017{\natexlab{b}})Mesaros, Heittola, and Virtanen}]{mesaros2017tut}
Annamaria Mesaros, Toni Heittola, and Tuomas Virtanen. 2017{\natexlab{b}}.
\newblock \href {https://doi.org/10.5281/zenodo.400515} {Tut acoustic scenes 2017, development dataset}.
\newblock Data set.

\bibitem[{Mesaros et~al.(2018)Mesaros, Heittola, and Virtanen}]{Mesaros2018_DCASE}
Annamaria Mesaros, Toni Heittola, and Tuomas Virtanen. 2018.
\newblock \href {https://arxiv.org/abs/1807.09840} {A multi-device dataset for urban acoustic scene classification}.
\newblock In \emph{Proceedings of the Detection and Classification of Acoustic Scenes and Events 2018 Workshop (DCASE2018)}, pages 9--13.

\bibitem[{OpenAI et~al.(2024)OpenAI, Achiam, and Others}]{openai2024gpt4technicalreport}
OpenAI, Josh Achiam, and Others. 2024.
\newblock \href {https://arxiv.org/abs/2303.08774} {Gpt-4 technical report}.
\newblock \emph{Preprint}, arXiv:2303.08774.

\bibitem[{Piczak()}]{piczak2015dataset}
Karol~J. Piczak.
\newblock \href {https://doi.org/10.1145/2733373.2806390} {{ESC}: {Dataset} for {Environmental Sound Classification}}.
\newblock In \emph{Proceedings of the 23rd {Annual ACM Conference} on {Multimedia}}, pages 1015--1018. {ACM Press}.

\bibitem[{Radford et~al.(2021)Radford, Kim, Hallacy, Ramesh, Goh, Agarwal, Sastry, Askell, Mishkin, Clark et~al.}]{radford2021learning}
Alec Radford, Jong~Wook Kim, Chris Hallacy, Aditya Ramesh, Gabriel Goh, Sandhini Agarwal, Girish Sastry, Amanda Askell, Pamela Mishkin, Jack Clark, et~al. 2021.
\newblock Learning transferable visual models from natural language supervision.
\newblock In \emph{International conference on machine learning}, pages 8748--8763. PMLR.

\bibitem[{Rafii et~al.(2017{\natexlab{a}})Rafii, Liutkus, St{\"o}ter, Mimilakis, and Bittner}]{musdb18}
Zafar Rafii, Antoine Liutkus, Fabian-Robert St{\"o}ter, Stylianos~Ioannis Mimilakis, and Rachel Bittner. 2017{\natexlab{a}}.
\newblock \href {https://doi.org/10.5281/zenodo.1117372} {The {MUSDB18} corpus for music separation}.

\bibitem[{Rafii et~al.(2017{\natexlab{b}})Rafii, Liutkus, St{\"o}ter, Mimilakis, and Bittner}]{rafii2017musdb18}
Zafar Rafii, Antoine Liutkus, Fabian-Robert St{\"o}ter, Stylianos~Ioannis Mimilakis, and Rachel Bittner. 2017{\natexlab{b}}.
\newblock The musdb18 corpus for music separation.

\bibitem[{Ruderman et~al.(2018)Ruderman, Rabinowitz, Morcos, and Zoran}]{ruderman2018pooling}
Avraham Ruderman, Neil~C Rabinowitz, Ari~S Morcos, and Daniel Zoran. 2018.
\newblock Pooling is neither necessary nor sufficient for appropriate deformation stability in cnns.
\newblock \emph{arXiv preprint arXiv:1804.04438}.

\bibitem[{Salamon et~al.(2014)Salamon, Jacoby, and Bello}]{Salamon:UrbanSound:ACMMM:14}
J.~Salamon, C.~Jacoby, and J.~P. Bello. 2014.
\newblock A dataset and taxonomy for urban sound research.
\newblock In \emph{22nd {ACM} International Conference on Multimedia (ACM-MM'14)}, pages 1041--1044, Orlando, FL, USA.

\bibitem[{Spadini(2019)}]{spadini2019sound}
Tito Spadini. 2019.
\newblock \href {https://doi.org/10.5281/zenodo.3519845} {Sound events for surveillance applications (1.0.0)}.
\newblock Data set.

\bibitem[{Springenberg et~al.(2014)Springenberg, Dosovitskiy, Brox, and Riedmiller}]{springenberg2014striving}
Jost~Tobias Springenberg, Alexey Dosovitskiy, Thomas Brox, and Martin Riedmiller. 2014.
\newblock Striving for simplicity: The all convolutional net.
\newblock \emph{arXiv preprint arXiv:1412.6806}.

\bibitem[{Tian et~al.(2014)Tian, Srinivasamurthy, Sandler, and Serra}]{bejingopera}
Mi~Tian, Ajay Srinivasamurthy, Mark Sandler, and Xavier Serra. 2014.
\newblock \href {https://doi.org/10.1109/ICASSP.2014.6853981} {A study of instrument-wise onset detection in beijing opera percussion ensembles}.
\newblock In \emph{2014 IEEE International Conference on Acoustics, Speech and Signal Processing (ICASSP)}, pages 2159--2163.

\bibitem[{Turian et~al.(2022)Turian, Shier, Khan, Raj, Schuller, Steinmetz, Malloy, Tzanetakis, Velarde, McNally et~al.}]{turian2022hear}
Joseph Turian, Jordie Shier, Humair~Raj Khan, Bhiksha Raj, Bj{\"o}rn~W Schuller, Christian~J Steinmetz, Colin Malloy, George Tzanetakis, Gissel Velarde, Kirk McNally, et~al. 2022.
\newblock Hear: Holistic evaluation of audio representations.
\newblock In \emph{NeurIPS 2021 Competitions and Demonstrations Track}, pages 125--145. PMLR.

\bibitem[{Tzanetakis et~al.(2001)Tzanetakis, Essl, and Cook}]{tzanetakis_essl_cook_2001}
George Tzanetakis, Georg Essl, and Perry Cook. 2001.
\newblock \href {http://ismir2001.ismir.net/pdf/tzanetakis.pdf} {Automatic musical genre classification of audio signals}.

\bibitem[{Wu et~al.(2022)Wu, Seetharaman, Kumar, and Bello}]{wu2022wav2clip}
Ho-Hsiang Wu, Prem Seetharaman, Kundan Kumar, and Juan~Pablo Bello. 2022.
\newblock Wav2clip: Learning robust audio representations from clip.
\newblock In \emph{ICASSP 2022-2022 IEEE International Conference on Acoustics, Speech and Signal Processing (ICASSP)}, pages 4563--4567. IEEE.

\bibitem[{Wu* et~al.(2023)Wu*, Chen*, Zhang*, Hui*, Berg-Kirkpatrick, and Dubnov}]{laionclap2023}
Yusong Wu*, Ke~Chen*, Tianyu Zhang*, Yuchen Hui*, Taylor Berg-Kirkpatrick, and Shlomo Dubnov. 2023.
\newblock Large-scale contrastive language-audio pretraining with feature fusion and keyword-to-caption augmentation.
\newblock In \emph{IEEE International Conference on Acoustics, Speech and Signal Processing, ICASSP}.

\bibitem[{Yang et~al.(2021)Yang, Chi, Chuang, Lai, Lakhotia, Lin, Liu, Shi, Chang, Lin et~al.}]{yang2021superb}
Shu-wen Yang, Po-Han Chi, Yung-Sung Chuang, Cheng-I~Jeff Lai, Kushal Lakhotia, Yist~Y Lin, Andy~T Liu, Jiatong Shi, Xuankai Chang, Guan-Ting Lin, et~al. 2021.
\newblock Superb: Speech processing universal performance benchmark.
\newblock \emph{arXiv preprint arXiv:2105.01051}.

\bibitem[{Zhang and Xiang(2023)}]{zhang2023decoupling}
Zihan Zhang and Xiang Xiang. 2023.
\newblock Decoupling maxlogit for out-of-distribution detection.
\newblock In \emph{Proceedings of the IEEE/CVF Conference on Computer Vision and Pattern Recognition}, pages 3388--3397.

\bibitem[{Zhou et~al.(2022)Zhou, Yang, Loy, and Liu}]{zhou2022learning}
Kaiyang Zhou, Jingkang Yang, Chen~Change Loy, and Ziwei Liu. 2022.
\newblock Learning to prompt for vision-language models.
\newblock \emph{International Journal of Computer Vision}, 130(9):2337--2348.

\end{thebibliography}
